\documentclass[a4paper,onecolumn,11pt,accepted=2022-12-05]{quantumarticle}
\pdfoutput=1
\usepackage[utf8]{inputenc}
\usepackage[english]{babel}
\usepackage[T1]{fontenc}
\usepackage{amsmath}
\usepackage{hyperref}
\usepackage[numbers,sort&compress]{natbib}

\usepackage{lipsum}

\usepackage{amsfonts,amssymb,tikz}
\usepackage{graphicx}
\usepackage{theorem}
\usepackage{color}
\usepackage{mathtools}
\usepackage{multirow} 

\usepackage{array}
\usetikzlibrary{quantikz}

\usepackage{dashbox}

\mathtoolsset{showonlyrefs}

\newtheorem{theorem}{Theorem}
\newtheorem{definition}{Definition}
\newtheorem{corollary}{Corollary}

\newtheorem{lemma}{Lemma}
\newtheorem{remark}{Remark}
\newtheorem{claim}{Claim}

\begin{document}

\title{Quantum algorithm for persistent Betti numbers and topological data analysis}

\author{Ryu Hayakawa}
\affiliation{Yukawa Institute for Theoretical Physics, Kyoto University, Kitashirakawa Oiwakecho, Sakyoku, Kyoto 606-8502, Japan}
\orcid{0000-0002-5403-1342}
\email{ryu.hayakawa@yukawa.kyoto-u.ac.jp}
\maketitle

\begin{abstract}
 Topological data analysis (TDA) is an emergent field of data analysis. The critical step of TDA is computing the persistent Betti numbers.
Existing classical algorithms for TDA are limited if we want to learn from high-dimensional topological features because the number of high-dimensional simplices grows exponentially in the size of the data. In the context of quantum computation, it has been previously shown that there exists an efficient quantum algorithm for estimating the Betti numbers even in high dimensions. However, the Betti numbers are less general than the persistent Betti numbers, and there have been no quantum algorithms that can estimate the persistent Betti numbers of arbitrary dimensions. 

This paper shows the first quantum algorithm that can estimate the ({\it normalized}) persistent Betti numbers of arbitrary dimensions. Our algorithm is efficient for simplicial complexes such as the Vietoris-Rips complex and demonstrates exponential speedup over the known classical algorithms.
\end{abstract}

\section{Introduction}
Processing large data has become a highly demanding task in modern society.
Topological data analysis (TDA) is an emergent field of information processing based on topological and geometric methods \cite{wasserman2018topological,edelsbrunner2010computational}. 
In particular, 
TDA based on persistent homology~\cite{edelsbrunner2000topological,zomorodian2005computing,edelsbrunner2008persistent,chazal2017introduction,otter2017roadmap}
has been gathering a lot of attention recently. 
TDA can
reveal the ``shape of data'' with multi-scale analysis of topological features. 
A crucial property of persistent homology, which is favorable in practical situations, is its stability against the noise concerning how the data is sampled \cite{cohen2007stability}.
There have been numerous applications of TDA, including
the protein structure
\cite{xia2014persistent},
the quantum many-body dynamics \cite{spitz2020finding},
the cosmic web
\cite{pranav2017topology}, and
the string theory 
\cite{cole2019topological},
to name a few.
Recently, combinations of persistent homology with supervised or unsupervised machine learning methods have also been actively studied \cite{pun2018persistent}. 

In the procedure of TDA, data is first converted into a nested sequence of topological objects. 
Such a topological object is called a {\it simplicial complex}, and a nested sequence of simplicial complexes is called {\it filtration}.
We can then study how topological invariants, such as the number of connected components, holes, voids, and their high-dimensional counterparts, change within the filtration using persistent homology. 
Typically, topological invariants that persist longer are considered to be more ``significant'' than those that persist shorter. 
The behavior of topological invariants is summarized as diagrams or functions in the form such as the persistent Barcodes \cite{ghrist2008barcodes} and the persistent Landscapes \cite{bubenik2015statistical}. 
Therefore, a typical procedure of TDA consists of first
constructing a family of topological objects from the given data and then computing the persistence of topological invariants, i.e. computing the persistent Betti numbers. (For the formal definition of persistent Betti numbers, see Section~\ref{sec:persistence}).

There are challenges in classical TDA that arise from the combinatorial nature of the topological treatment of data.  
Assume that the input data is composed of $n$ points. 
Then, the number of possible $q$-dimensional elements (i.e. $q$-simplices) that can contribute to topology is ${n \choose q+1}$. 
(We say a simplex is $q$-dimensional if it is composed of $q+1$ vertices.) 
The known classical algorithms for persistent homology work in polynomial time {\it in the number of simplices} \cite{zomorodian2005computing,memoli2020persistent}. 
Classical algorithms work in polynomial time in $n$ for constant dimensional cases because the number of constant-dimensional simplices is at most polynomial in $n$. 
On the other hand, there can be exponentially many simplices for super constant dimensions. 
As a result,
performing a high-dimensional analysis of TDA is a difficult task for classical computers. 
There are many studies to overcome this issue such as 
the way of reducing the size of simplicial complexes while keeping their homology unchanged  \cite{barmak2012strong}. 
However, such techniques are not enough to overcome the exponential growth of simplices in high dimensions.

\paragraph{Existing quantum approaches for TDA.}
Data learning tasks are also actively studied in the field of quantum computation. 
In order to investigate the quantum computational approaches for TDA, 
Lloyd, Garnerone, and Zanardi \cite{lloyd2016quantum} 
proposed a quantum algorithm (LGZ algorithm) for estimating the (normalized)\footnote{``Normalized'' means that Betti number is divided by the total number of $q$-dimensional elements of the simplicial complex.} {\it Betti numbers}. 
The $q$-th Betti number represents the number of ``$q$-dimensional holes'' of a simplicial complex. 
By estimating the Betti numbers for simplicial complexes that arise from each step of the filtration, the LGZ algorithm can be used to learn the topological structures of data. 
A function that represents how the Betti numbers vary within the filtration is called a {\it Betti function} or a {\it Betti curve} \cite{chazal2017introduction}.
Therefore, the LGZ algorithm can approximately recover the Betti curve of arbitrary dimensions.
The LGZ algorithm demonstrates an exponential speedup over the known best classical algorithm for estimating the high-dimensional Betti numbers. 
A proof of principle experiment of the LGZ algorithm has also been done \cite{huang2018demonstration}. 

Estimating the Betti numbers for each element of the filtration is not enough to fully perform TDA based on persistent homology.
To perform such a full TDA, it is crucial to estimate the {\it persistent Betti numbers}. 
The $q$-th $(t,s)$-persistent Betti number is
the number of $q$-dimensional holes that exist at scale parameter $t$ that are still alive at $s$. (For a formal definition, see Section~\ref{sec:persistence}.)
Because the $(t,t)$-persistent Betti number is just the usual Betti number, estimating the persistent Betti numbers is a more general task than estimating the Betti numbers. 
It is shown in \cite{meijer2019clustering,neumann2019limitations} that the LGZ algorithm can be used to estimate the persistent Betti numbers of the zeroth dimension but fail in higher dimensions in general. 
The TDA of the zeroth dimension is not the region where we can exploit the benefit of quantum computation 
because the classical algorithms run in polynomial time for constant dimensions. 
Therefore, constructing a quantum algorithm for persistent Betti numbers for arbitrary dimensions has been a significant open problem for the quantum TDA.

\paragraph{Our result.}
We present a first quantum algorithm for estimating the persistent Betti numbers of arbitrary dimensions. 
We provide error and complexity analysis and the condition for the algorithm to succeed. 
The formal statement can be seen in Theorem~\ref{them:main} in Section~\ref{sec:main}. 
We resolve the open problem of estimating the persistent Betti numbers by estimating the spectral density of the ground states of a certain positive-semidefinite hermitian operator using quantum computation. 
The positive-semidefinite operator is called {\it persistent combinatorial Laplacian}~\cite{wang2020persistent,memoli2020persistent}. 
(The persistent combinatorial Laplacian is different from the combinatorial Laplacian, which is used in the LGZ algorithm.)
Estimating the persistent Betti numbers allows one to recover the topological summary of data such as the persistent Barcodes.

\paragraph{Comparison with the classical TDA.}
The best known classical algorithm for computing the persistent Betti numbers of a simplicial pair $K\hookrightarrow L$  \cite{memoli2020persistent} runs in time $\mathcal{O}((n^L_q)^\omega)$ for $\omega < 2.373$, where $n^L_q$ is the number of $q$-simplices of $L$ (assuming $n^L_q=\mathcal{O}(n^L_{q+1})$, which holds in many practical cases such as for Vietoris-Rips or \v{C}ech complexes). 
For an approximation task, the best-known classical algorithms for estimating the spectral densities of matrices 
\cite{cheung2013fast,di2016efficient,lin2016approximating}
runs time linearly with the number of non-zero elements of the matrices. 
The known classical algorithm for computing the matrix representation of the persistent Laplacian ~\cite{memoli2020persistent} takes $\mathcal{O}((n^L_q)^\omega+n^L_{q+1})$ time.
Therefore, there are no known classical algorithms for computing or approximating the persistent Betti numbers efficiently when $n^L_{q}$ is exponential in the number of vertices $n$. 

{We present our quantum algorithm using the membership oracle for simplicial complexes. 
Therefore, it is necessary that we can efficiently construct the membership function so that we can efficiently run the quantum algorithm. }
Indeed, we can efficiently implement the membership functions for the construction of simplicial complexes such as the Vietoris-Rips complex and the lazy witness complex, as we see in Appendix~\ref{section:practical}.
Therefore, our quantum algorithm demonstrates exponential speedup over the best-known classical algorithm for instances of problems that satisfy the promises of Theorem~\ref{them:main}.

Nevertheless, the drawback of our result is that we can only estimate the normalized version of the persistent Betti numbers. It is not known whether the estimation of such normalized persistent Betti number is useful or not. 
It is an important future direction to understand whether there are practical instances of problems that satisfy the conditions of Theorem~\ref{them:main} and whether the estimation of the normalized persistent Betti number is useful or not.

\paragraph{Related Work.}

The LGZ algorithm~\cite{lloyd2016quantum} estimates the Betti numbers using the quantum phase estimation algorithm. 
Based on our quantum algorithm for the persistent Betti numbers,
we can immediately provide a quantum algorithm for estimating the normalized Betti numbers based on the block-encoding and QSVT by slightly modifying our algorithm (because Betti numbers are special cases of persistent Betti numbers).
The gate complexity of implementing a quantum algorithm which outputs $1$ with probability $\tilde{p}_1$ s.t.
\begin{equation}
    |\tilde{p}_1 - {\beta^K_q}/{n^K_q}|\leq \epsilon
\end{equation}
is $\mathcal{O}\left(poly(n)\log{(1/\epsilon)}\right)$, which is an exponential improvement in terms of the error parameter compared to the LGZ algorithm. (Nonetheless, this exponential improvement does not actually improve the complexity in terms of error due to the sampling error.)

Recently, a NISQ algorithm for the Betti numbers is presented in \cite{ubaru2021quantum}. Their algorithm is based on the following observations: (a) a boundary map can be represented as a sum of Pauli operators; 
(b) Instead of using Grover's quantum search to implement the uniform mixture of the $q$-simplex states, a quantum rejection sampling can be used; 
(c) a stochastic rank estimation can be used to estimate the Betti numbers instead of the quantum phase estimation. 
Based on these observations, their quantum algorithm achieves a depth complexity that is linear in $n$ at the cost of the success probability $1/d^K_q$ of the rejection sampling. 
It is an open problem to construct a quantum algorithm for the persistent Betti numbers in a way that is preferable for NISQ devices.

\paragraph{Organization.}
The rest of the paper is organized as follows. In Section \ref{sec:persistence}, we give the preliminaries on persistent homology and TDA. 
In Section \ref{sec:qsvt}, we review the block-encoding and quantum singular value transformation. In Section 
{\ref{sec:main}, we provide a quantum algorithm for estimating the persistent Betti numbers, which is our main result. 
In Section~\ref{section:statepreparation} and Section~\ref{section:projector}, we present the implementation of some unitaries which are required to conclude the proof of the main result. }

\section{Preliminaries on persistent homology}
\label{sec:persistence}

In this section, we introduce persistent homology for simplicial complexes.
We first introduce some necessary terminologies such as the simplicial complex, the filtration, the Betti numbers, and the persistent Betti numbers. 
Then we introduce the recent result of \cite{memoli2020persistent} that the persistent Betti number can be calculated as the nullity of the so-called persistent Laplacian operator (Theorem~\ref{them:1}) and the persistent Laplacian can be implemented using the Schur complement (Theorem~\ref{theorem:schur}).
Table~\ref{table:notations} may be also helpful to overview the notations.

An abstract simplicial complex $K$ over a finite ordered set $V$ of vertices is a collection of subsets of $V$ such that for any $\sigma\in K$, if $\tau \subseteq \sigma$ then $\tau \in K$.
An element $\sigma\in K$ is called a $q$-simplex if $|\sigma|=q+1$.
For example, a 0-simplex is called a vertex, and a 1-simplex is called a line. 
We denote the set of $q$-simplices of $K$ by $S^K_q$.

An oriented simplex
is a simplex in $K$ with a fixed ordering of vertices inherited from the ordering of $V$. 
We denote an oriented simplex of $\sigma\in K$ as $\tilde{\sigma}$.
Let $\Bar{S}^K_q\coloneqq \{\tilde{\sigma}:\sigma\in S^K_q\}$ be a set of oriented $q$-simplices of $K$. 
The $q$-th chain group $C^K_q$ of $K$ is the vector space over ${\mathbb R}$ with basis $\Bar{S}^K_q$ \footnote{
We work with chain groups with coefficients $\mathbb{R}$, 
which is suitable for our quantum algorithm that relies on the Hodge theory and the persistent Laplacian. 
The classical algorithms for persistent homology of \cite{zomorodian2005computing} work for field coefficients such as $\mathbb{Z}_2$. 
The Betti numbers of homology groups over $\mathbb{R}$ coefficients do not capture the torsion \cite{lim2015hodge}. 
Nonetheless, this seems to be enough for many of the situations in TDA.
}. 
Let $n^K_q\coloneqq {\rm dim}(C^K_q)=|S^K_q|$ be the dimension of $C^K_q$.

The boundary map $\partial^K_q:C^K_q \rightarrow C^K_{q-1}$ 
is defined as 
\begin{equation}
\label{eq:boundary}
    \partial^K_q([v_0,...,v_q]) \coloneqq 
    \sum_{i=0}^q
    (-1)^i[v_0,...,\hat{v}_i,...,v_q]
\end{equation}
for each $\tilde{\sigma}=[v_0,...,v_q]$, where the hat indicates that the vertex $v_i$ is omitted. 
Then, the $q$-th homology group of $K$ is
\begin{equation}
    H_q(K) \coloneqq {\rm ker}(\partial^K_q)/ {\rm im}(\partial ^K_{q+1}),
\end{equation}
and the $q$-th Betti number of $K$ is 
\begin{equation}
    \beta^K_q \coloneqq {\rm rank}(H_q(K)).
\end{equation}
The $q$-th combinatorial Laplacian $\Delta^K_q:C^K_q\rightarrow C^K_q$ is defined as follows:
\begin{equation}
    \Delta^K_q \coloneqq 
    \underbrace{
    \partial^K_{q+1} \circ (\partial^K_{q+1})^{*}
    }_{=:\Delta^K_{q,{\rm up}}}
    +
    \underbrace{
    (\partial^K_q)^{*}\circ \partial^K_q
    }_{=:\Delta^K_{q,{\rm down}}},
\end{equation}
where $(\partial^K_q)^{*}:C^K_{q-1}\rightarrow C^K_q$ is the adjoint of $\partial^K_q$ under the inner product $\langle \cdot,\cdot \rangle_{C^K_q}$ such that $\Bar{S}^K_q$ is an orthonormal basis of $C^K_q$. 
We have introduced $\Delta^K_{q,{\rm up}}, \Delta^K_{q,{\rm down}}$ for the convenience of notation.  
The Betti number can be calculated as the nullity of the combinatorial Laplacian \cite{friedman1998computing}:
\begin{equation}
  \label{eq:betti}
   \beta^K_q = {\rm nullity}(\Delta^K_q)
   = \#\ {\rm of\ zero\ eigenvalues\ of\ } \Delta^K_q.
\end{equation}
This relationship comes from the following elementary result of Hodge theory:

\begin{claim}[\cite{lim2015hodge}]
\label{claim:hodge}
Let $A\in \mathbb{R}^{m\times n}$ and $B \in \mathbb{R}^{n\times p}$, and assume $AB=0$. 
Then we have 
\begin{equation}
    \rm{ker}(A)/\rm{im}(B) \cong \rm{ker}(BB^*+A^*A).
\end{equation}
\end{claim}
Eq.~\eqref{eq:betti} follows using this claim observing that $\partial^K_q \partial^K_{q+1}=0$.

\subsection{Filtration and Persistent homology}

A sequence of nested subcomplexes of $F$
\begin{equation}
    \emptyset= F_0 \subseteq F_1 \subseteq ... \subseteq F_m = F
\end{equation}
is called filtration. 
Similarly, we say a pair of simplicial complexes $K$ and $L$ is a simplicial pair if $K$ and $L$ are the simplicial complexes over the same ordered set $V$ such that $K\subseteq L$. We denote a simplicial pair by $K \hookrightarrow L$. 
For a simplicial pair $K\hookrightarrow L$, 
we choose the ordering $\{\sigma^L_q(i)\}_{i=1}^{n^L_q}$ such that $\{\sigma^L_q(i)\}_{i=1}^{n^K_q}=\{\sigma^K_q(i)\}_{i=1}^{n^K_q}$.
We obtain a map between the $q$-th homology group of $K$ and $L$ induced by the inclusion:
\begin{equation}
    f^{K,L}_q : H_q(K)\rightarrow H_q(L). 
\end{equation}
The $q$-th persistent homology group $H^{K,L}_q$ is defined as the image of the map. 
The $q$-th persistent Betti number $\beta^{K,L}_q$ is defined as the rank of the map. 
The persistent homology group consists of the homology group of $K$ that are still alive at $L$. That is, 
\begin{equation}
    H^{K,L}_q= {\rm ker}(\partial^K_q)/ 
    \left( {\rm im}(\partial^L_{q+1})\cap{\rm ker}(\partial^K_q)\right).
\end{equation}
Note that the difference of $q$-th Betti numbers of $K$ and $L$ ($\beta_q^L-\beta_q^K$) does not reveal the same information as $\beta^{K,L}_q$ because $\beta^{K,L}_q$ is the number of topological invariants that exist at $K$ that are still alive at $L$. 

\subsection{Persistent Laplacian}
The persistent Laplacian is first introduced in \cite{wang2020persistent}. 
Later, the property and implementations of the persistent Laplacian are further studied in \cite{memoli2020persistent}. Here, we introduce the $q$-th persistent Laplacian for a persistent pair $K \hookrightarrow L$. 
Consider the subspace of $C^L_q$
\begin{equation}
    C^{L,K}_q \coloneqq \{
    c \in C^L_q : \partial^L_q(c) \in C^K_{q-1}
    \}.
\end{equation}
Then, let $\partial^{L,K}_q$ be the restriction of the boundary operator $\partial^L_q$ to $C^{L,K}_q$ so that the image of $\partial^{L,K}_q$ is contained in $C^K_{q-1}$. 
Then, the $q$-th persistent Laplacian is defined as

\begin{equation}
 \Delta_q^{K,L}\coloneqq 
 \underbrace{
 \partial_{q+1}^{L,K}\circ (\partial_{q+1}^{L,K})^{*}
 }_{=:\Delta^{K,L}_{q,{\rm up}}}
 +
 \underbrace{
  (\partial_{q}^K)^{*}\circ \partial_{q}^K
 }_{=\Delta^{K}_{q,{\rm down}}}.
\end{equation}

It is shown in \cite{memoli2020persistent} that the nullity of the $q$-th persistent Laplacian equals to the $q$-th persistent Betti numbers:

\begin{theorem}[\cite{memoli2020persistent}]
\label{them:1}
For each integer $q\geq 0$, 
$\beta_q^{K,L}={\rm nullity}(\Delta_q^{K,L})$.
\end{theorem}
We provide a proof in Appendix~\ref{app:proof} for completeness.

In addition to the persistent Betti numbers from the nullity of the persistent Laplacian, it is suggested in \cite{wang2020persistent} that the spectral property of the persistent Laplacian beyond the zero eigenvalues would provide useful information about the data.

\paragraph{Matrix representation.}
We can give a matrix representation of  $\partial^K_q$ with respect to the canonical basis $\{\sigma^K_q(i)\}_{i\in [n^K_q]}$ of $C^K_q$ and $\{\sigma^K_{q-1}(i)\}_{i\in [n^K_{q-1}]}$ of $C^K_{q-1}$
according to eq.~\eqref{eq:boundary}.
The matrix representation $B^K_q$ of $\partial^K_q$ is $n^K_{q-1}\times n^K_{q}$ matrix with entries from $\{-1,0,+1\}$. 
For example, if the canonical basis of $C^K_1$ is $\{01,02,03,12\}$ and the canonical basis of $C^K_0$ is $\{0,1,2,3\}$, the boundary matrix is represented as
\begin{equation}
    B^K_1 = 
\begin{array}{c c} &
\begin{array}{c c c c} 01 & 02 &03 & 12 \\
\end{array}
\\
\begin{array}{c c c c}
0 \\
1\\
2\\
3
\end{array}
&
\left[
\begin{array}{c c c c}
-1 & -1 & -1 &0 \\
1 & 0 & 0 &-1\\
0 & 1 & 0&1\\
0 & 0 & 1&0
\end{array}
\right].
\end{array}
\end{equation}
$B^K_q$ is $(q+1)$-column sparse by definition. 
The matrix representation of $\left(\partial^K_q\right)^*$ is simply $(B^K_q)^\mathsf{T}$. 
Choose any basis of $C^{L,K}_{q+1}$ represented by a column matrix $Z\in \mathbb{R}^{n^L_{q+1}\times n^{L,K}_{q+1}}$ and let $B^{L,K}_{q+1}$ be the corresponding matrix representation of $\partial^{L,K}_{q+1}$.
It is shown in \cite{memoli2020persistent} that the matrix representation of $(\partial^{L,K}_{q+1})^*$ is $\left(Z^\mathsf{T}Z\right)^{-1}\left(B^{L,K}_{q+1}\right)^\mathsf{T}$. 
Then the matrix representation of the persistent Laplacian is 
\begin{equation}
    \label{eq:matrixrep}
    \Delta^{K,L}_q
    =B^{L,K}_{q+1}\left(Z^\mathsf{T}Z\right)^{-1}\left(B^{L,K}_{q+1}\right)^\mathsf{T}
    +
    \left(B^K_q\right)^\mathsf{T}B^K_q.
\end{equation}

In \cite{memoli2020persistent}, 
the authors have given two algorithms for computing the matrix representation of $\Delta_q^{K,L}$. 
We adopt the second algorithm which uses the Schur complement. 
The Schur complement is defined as follows.

\begin{definition}[Schur complement]
Let $M\in \mathbb{R}^{n\times n}$ be a block matrix $ M =
\begin{pmatrix}
A & B \\
C & D \\
\end{pmatrix} $ where $D\in \mathbb{R}^{d\times d}$ is a block matrix.
The Schur complement of $D$ in $M$ is 
\begin{equation}
    M/D \coloneqq 
    A-BD^+ C,
\end{equation}
where $D^+$ is the Moore-Penrose pseudo-inverse of $D$.
Similarly, for $\emptyset \neq I \subseteq [n]$ and $\emptyset \neq J \subseteq [n]$, we denote the submatrix of $M$ consisting of rows and columns of indices of $I$ and $J$ by $M(I,J)$. Then, the Schur complement of $M(I,I)$ in $M$ is 
\begin{equation}
    M/M(I,I) \coloneqq
    M([n]\backslash I,[n]\backslash I)-
    M([n]\backslash I,I)M(I,I)^+ M(I,[n]\backslash I),
\end{equation}
where 
 $[n]\backslash I\coloneqq \{a\in[n]\ |\ a\notin I\}$
 and 
 $M(I,J)$ is the submatrix of $M$ that consists of rows and columns of $M$ indexed by $I$ and $J$, respectively.
\end{definition}
It is shown in \cite{memoli2020persistent} that $ \Delta^{K,L}_{q,{\rm up}}$ can be implemented using the Schur complement: 
\begin{theorem}[\cite{memoli2020persistent}]
\label{theorem:schur}
Let $K\hookrightarrow L$ be a simplicial pair. 
Assume that $n^K_q<n^L_q$ and let $I^L_K\coloneqq [n^L_q]\backslash [n^K_q]$. 
Then, 
\begin{equation}
    \Delta^{K,L}_{q,{\rm up}} = \Delta^{L}_{q,{\rm up}}/\Delta^{L}_{q,{\rm up}}(I^L_K,I^L_K).
\end{equation}
\end{theorem}

{
\begin{remark}
Finding the orthonormal basis of $C^K_q$ is non-trivial and therefore building $B^K_q$ is also non-trivial. If the basis can be found (by matrix reduction, see~\cite{memoli2020persistent} for detail), we can find $Z$ and $B_q^{L,K}$, and compute $    \Delta^{K,L}_q
=B^{L,K}_{q+1}\left(Z^\mathsf{T}Z\right)^{-1}\left(B^{L,K}_{q+1}\right)^\mathsf{T}
    +
\left(B^K_q\right)^\mathsf{T}B^K_q$
.
For the quantum algorithm we present in this paper, we do not need to classically compute the matrix representation of $\Delta^{K,L}_q$ because we can bypass the necessity of computing the matrix representation of $\Delta^{K,L}_q$
by building the block-encoding of $\Delta^{K,L}_q$ using the quantum membership functions assisted by the method of~\cite{memoli2020persistent}, which uses the Schur complement.
\end{remark}
}

\subsection{Correspondence with Quantum computation}

We can use the Hilbert space of $n$-qubits to represent a chain complex of a simplicial complex over $n$-vertices.
We can correspond a $q$-simplex to an $n$-bit string of Hamming weight $q+1$, 
where the indices of 1s correspond to the indices of vertices of a simplex. 
We denote $|\tilde{\sigma}_q\rangle$ where $\tilde{\sigma}_q$ is an $n$-bit string with Hamming weight $q+1$. 
We denote the ${n \choose q+1}$ dimensional Hilbert space in 
${\mathbb C}^{2^n}$ spanned by computational basis states of Hamming weight $q+1$ by ${\mathcal W}_q$. 
Let ${\mathcal H}^K_q$ be the subspace of ${\mathcal W}_q$ spanned by $\{|\tilde{\sigma}_q\rangle : \tilde{\sigma}_q \in \bar{S}^K_q \}$.
We call $\{|\tilde{\sigma}_q\rangle : \tilde{\sigma}_q \in \bar{S}^K_q \}$ the canonical basis of ${\mathcal H}^K_q$. 
The superposition of quantum states $\{|\tilde{\sigma}_q\rangle : \tilde{\sigma}_q \in \bar{S}^K_q \}$ is the quantum state representation of the $q$-chain. 
The boundary operator maps as $\partial^K_q: {\mathcal H}^K_q \rightarrow {\mathcal H}^K_{q-1}$. 
A summary of notations is given in Table~\ref{table:notations}.
\renewcommand{\arraystretch}{1.3}
\begin{table}[thbp]
  \caption{Summary of notations}
  \label{table:notations}
  \centering
  \begin{tabular}{cl}
    \hline
    $K\hookrightarrow L$ & A simplicial pair\\ \hline
    $n$ & The number of vertices \\ \hline
    $\bar{S}^K_q$ & The set of oriented $q$-simplices in $K$\\\hline
    $C^K_q$ & The $q$-th chain group of $K$ with  coefficients $\mathbb{R}$\\ \hline
    $n^K_q$ & ${\rm dim} (C^K_q) = |S^K_q|$\\\hline
    $\{\sigma^K_q(i)\}_{i\in [n^K_q]}$ & canonical basis of $C^K_q$ \\\hline
    ${\mathcal W}_q$ & Subspace of ${\mathbb C}^{2^n}$ spanned by Hamming weight $q+1$ states\\ \hline
    %$\{|\sigma^K_q(i)\rangle\}_{i\in [n^K_q]}$ & canonical basis of ${\mathcal H}^K_q$ \\ \hline
    ${\mathcal H}^K_q$ & Subspace of ${\mathcal W}_q$ spanned by  $\{|\sigma^K_q(i)\rangle\}_{i\in [n^K_q]}$  \\\hline
    $\{|\psi_q(i)\rangle\}_{i\in [n^K_q]}$ &  Eigenbasis of $\Delta^{K,L}_q$ (in ${\mathcal H}^K_q$) with eigenvalues $\lambda_1\leq \lambda_2\leq...\leq\lambda_{n^K_q}$ \\\hline
  \end{tabular}
\end{table}
\renewcommand{\arraystretch}{1.0}

\section{Block-encoding and quantum singular value transformation}
\label{sec:qsvt}

In this section, we review the block-encoding of quantum operators and the quantum singular value transformation (QSVT). 
We refer the reader to \cite{gilyen2019quantum,martyn2021grand} for more detail. 
We use $\|\cdot\|$ as the spectral norm 
and $\|\cdot\|_\diamond$ as the diamond norm. 
The diamond norm is defined as $\|\Lambda\|_\diamond\coloneqq \underset{\rho}{\rm{max}} \|(\Lambda\otimes \mathcal{I})(\rho)\|_1$, where $\|\cdot\|_1$ is the trace norm.
We also use the notation of $\tilde{\mathcal{O}}(\cdot)$ which omits  $poly(\log n)$ and $poly(\log(\log\frac{1}{\epsilon}))$ factors.

\subsection{Block-encoding and QSVT}
Let $A=\tilde{\Pi}U\Pi$, where $\tilde{\Pi}$, $\Pi$  are orthogonal projectors and $U$ is a unitary.
We call such $U$ a projected unitary encoding of $A$.
Especially,
a block-encoding of a quantum operator is defined as follows:
\begin{definition}[Block-encoding~\cite{gilyen2019quantum}]
Suppose that $A$ is an $s$-qubit operator, $\alpha,\epsilon\in{\mathbb R}_+$, and $a\in{\mathbb N}$.
We say that the $(s+a)$-qubit unitary $U$ is an $(\alpha,a,\epsilon)$-block-encoding of A, if
\begin{equation}
    \|A-\alpha(\langle 0|^{\otimes a}\otimes I^{\otimes s})U(| 0\rangle^{\otimes a}\otimes I^{\otimes s})\|\leq \epsilon.
\end{equation}
Here, $I$ is the $1$-qubit identity operator.
\end{definition}
Such encoding is called a block-encoding because $U$ approximates the block matrix
\begin{equation}
\left[
\begin{array}{cc}
A/\alpha & . \\
. & . \\
\end{array}
\right].
\end{equation}

It is shown in \cite{gilyen2019quantum} how to implement a unitary which realizes polynomial transformation of the singular values of the block-encoded matrices. 
Let $A=\tilde{\Pi}U\Pi$, $d\coloneqq {\rm rank}(\Pi)$, $\tilde{d}\coloneqq {\rm rank}(\tilde{\Pi})$ and $d_{\rm min}\coloneqq {\rm min}(d,\tilde{d})$. 
Then by the singular value decomposition, 
there exists orthonormal bases $\{\ket{\psi_i}:i\in[d]\}$ and $\{|\tilde{\psi}_i\rangle:i\in[\tilde{d}]\}$ of ${\rm img}(\Pi)$ and ${\rm img}(\tilde{\Pi})$ and $a_i\in {\mathbb{R}}_0^+$ such that
\begin{equation}
    A=\sum_{i=1}^{d_{\rm min}} a_i |\tilde{\psi_i}\rangle\bra{\psi_i},
\end{equation}
where $a_1\geq a_2\geq \cdots \geq a_{d_{\rm min}}$.
Then, the quantum singular value transformation (QSVT) of $A$ is defined 
for odd or even polynomial function $f:{\mathbb R}\rightarrow {\mathbb C}$
as
\begin{equation}
    f^{(\rm{SV})}(A)\coloneqq
    \left\{
    \begin{array}{l}
       \sum_{i=1}^{d_{\rm min}} f(a_i)|\tilde{\psi_i}\rangle\bra{\psi_i},\ {\rm if}\ f\ {\rm is\ odd},\\
       \sum_{i=1}^d f(a_i)|\psi_i\rangle\bra{\psi_i},\ {\rm if}\ f\ {\rm is\ even.}
    \end{array}
    \right.
\end{equation}
The following is shown in \cite{gilyen2019quantum}.
This is a slightly modified version of Corollary 18 of  \cite{gilyen2019quantum}. 

\begin{theorem}[Singular Value Transformation by real polynomial]
\label{theorem:qsvt}
Let ${\mathcal H}_U$ be a finite dimensional Hilbert space and 
let $U$, $\Pi$, $\tilde{\Pi}$ be linear operators that map ${\mathcal H}_U\rightarrow {\mathcal H}_U$ such that $U$ is a unitary, and $\Pi, \tilde{\Pi}$ are orthogonal projectors.
Suppose $f\in {\mathbb R}[x]$ be a degree-$d$ odd or even polynomial satisfying $|f(x)|\leq 1$ for all $x\in [-1,1]$.
Then, there exists an ${\mathcal{O}(poly(d,\log(1/\delta)))}$-time classical algorithm that outputs 
the classical description of a quantum circuit $U_{\Phi^+}$ and $U_{\Phi^+}'$ such that
\begin{equation}
    P_{\Re}^{(\rm{SV})}(\tilde{\Pi}U\Pi)=
    \begin{cases}
    \big(\langle 0|\otimes\tilde{\Pi}\big)
    U_{\Phi^+}
    \big(| 0\rangle\otimes\Pi\big),\ {\rm if}\ d\ {\rm is\ odd},\\
    \big(\langle 0|\otimes\tilde{\Pi}\big)
    U'_{\Phi^+}
    \big(| 0\rangle\otimes\Pi\big),\ {\rm if}\ d\ {\rm is\ even}
    \end{cases}
\end{equation}
and 
\begin{equation}
    \left\|f^{(\rm{SV})}(\tilde{\Pi}U\Pi)-P_{\Re}^{(\rm{SV})}(\tilde{\Pi}U\Pi)\right\|\leq \delta.
\end{equation}
Moreover, 
$U_{\Phi^+}$ and $U'_{\Phi^+}$ 
can be implemented using ${\mathcal O}(d)$ use of $U$, $U^\dagger$, ${\rm C}_\Pi {\rm NOT}$, ${\rm C}_{\tilde{\Pi}} {\rm NOT}$ and other single qubit gates, where ${\rm C}_\Pi {\rm NOT}$ is defined as 
\begin{equation}
 \label{eq:cpinot}
    {\rm C}_\Pi {\rm NOT} \coloneqq 
    X\otimes \Pi + I\otimes (I-\Pi).
\end{equation}
\end{theorem}
We give a proof in
Appendix \ref{app:proof}.
The corollary follows when $U$ is a block-encoding of $A$.

\begin{corollary}[QSVT by real polynomial for block-encoded operator]
\label{corol:qsvtreal}

Let $U$ be an \\ $(\alpha,a,0)$-block-encoding of an $n$-qubit operator $A$. Let $f\in {\mathbb R}[x]$ be a degree-$d$ odd or even polynomial satisfying $|f(x)|\leq 1$ for all $x\in [-1,1]$.
Then, there exists an ${\mathcal{O}(poly(d,\log(1/\delta)))}$-time classical algorithm that outputs a description of a quantum circuit $U_{\Phi^+}$ s.t. $U_{\Phi^+}$ is a $(1,a+1,\delta)$-block encoding of $f^{(\rm{SV})}(A/\alpha)$.
Moreover, $U_{\Phi^+}$ can be implemented with $\mathcal{O}(d)$ use of $U$, $U^\dagger$ and $\mathcal{O}\left(ad\right)$ other elementary quantum gates. 
\end{corollary}

\subsection{QSVT with some useful polynomials}

In this paper, we use QSVT with three different polynomials. 
First, we introduce the fixed-point amplitude amplification using polynomial approximations of the sign function (Lemma 25 of \cite{gilyen2019quantum}):

\begin{theorem}[Fixed-point amplitude amplification~\cite{gilyen2019quantum}]
\label{theorem:fixedpointamp}
Let $U$ be a unitary and $\Pi$ be an orthogonal projector such that $\Pi U |\psi_0\rangle=a|\psi_G\rangle$, and $a>\delta > 0$. 
There is a unitary circuit $\tilde{U}$ such that 
{
\begin{equation}
    \left\|
 \tilde{U}|\psi_0\rangle
    -
    |\psi_G\rangle
    \right\|
    \leq \epsilon,
\end{equation}
}
which uses a single ancilla qubit and consists of ${\mathcal O}(\log{(1/\epsilon)}/\delta)$-use of $U$, $U^\dagger$, $C_\Pi NOT$, 
$C_{|\psi_0\rangle \langle \psi_0 |}NOT$ and other quantum gates.
\end{theorem}
We use this theorem for state preparation of our algorithm in Section \ref{section:statepreparation}.

Next, we introduce the implementation of the Moore-Penrose pseudo-inverse. 
Suppose $\tilde{\Pi}U\Pi=A$ and
$A=W\Sigma V^\dagger$ is a singular value decomposition, where $\Sigma$ is a diagonal matrix with non-negative and non-increasing entries. 
Then the pseudo-inverse of $A$ is defined as
$A^+\coloneqq V\Sigma^+W^\dagger$,
where $\Sigma^+$ contains the inverse of the diagonal elements of $\Sigma$ except for $0$, which remains $0$. 
We can implement the pseudo-inverse using the polynomial approximation of $1/x$~\cite{gilyen2019quantum}:

\begin{theorem}[Implementing the Moore-Penrose pseudo-inverse]
\label{theorem:pseudoinverse}
Suppose that $A$ is an $n$-qubit positive-semidefinite hermitian operator and
$U$ is an $(\alpha,a,0)$-block-encoding of $A$. 
Assume that the smallest non-zero eigenvalue of $A$ is $\lambda_{\rm{min}}$ and let  $0< 1/\kappa <\lambda_{\rm{min}}/\alpha$ and $0<\epsilon<1/2$.
Then there is a $d={\mathcal O}(\kappa \log{(\frac{\kappa}{\alpha\epsilon}}))$ and an efficiently computable $\Phi\in {\mathbb R}^d$ such that 
$U_{\Phi^+}$ is a $(2\kappa/\alpha,a+1,\epsilon)$-block-encoding of $A^+$.
\end{theorem}
This is a slight modification of Theorem 41 of \cite{gilyen2019quantum} 
for the block-encoding of the positive-semidefinite operator. 
A proof is given in Appendix \ref{app:proof}.
We also use QSVT with a polynomial approximation of the rectangle function in Section~\ref{imp:4} in order to implement the projector.

\subsection{Block-measurement}
\label{subsec:review:blockmeasurement}

Finally, we review the block-measurement which is introduced in \cite{rall2021faster}.
We can implement a quantum channel that is close to the following map
\begin{equation}
    \label{eq:BMmap}
    |0\rangle\otimes|\psi\rangle
    \rightarrow
    |1\rangle\otimes\Pi |\psi\rangle 
    +|0\rangle\otimes (I-\Pi)|\psi\rangle
\end{equation}
using the approximate block-encoding of $\Pi$.
Let $U_{\Pi}$ be the {\it exact} block-encoding of $\Pi$ with $m$ ancilla qubits. 
Then, consider the following circuit $V$.
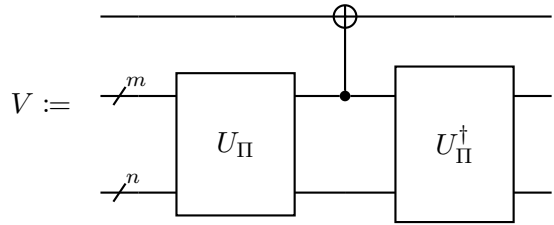
\begin{figure}[h]
\centering
$V\coloneqq$
\begin{quantikz}
&\qw&\qw & \targ{} &  \qw &\qw \\
&\qwbundle{m} &\gate[wires=2]{\ \ \ U_{\Pi}\ \ \ } & \ctrl{} \vqw{-1} & \gate[wires=2]{\ \ \ U^\dagger_{\Pi}\ \ \ } &\qw \\
&\qwbundle{n}& & \qw & &\qw
\end{quantikz}
\caption{The quantum circuit for block-measurement.}
\label{fig:blockmeasurement}
\end{figure}\\
The CNOT gate refers to $X\otimes \ket{0^m}\bra{0^m}+I\otimes (I_m-\ket{0^m}\bra{0^m})$ where $X$ is the Pauli-$X$ matrix.
It can be seen that $V$ satisfies 
\begin{equation}
    (I\otimes\bra{0^m}\otimes I_n) V (\ket{0}\otimes \ket{0^m}\otimes I_n)
    =
    \ket{1}\otimes\Pi+\ket{0}\otimes (I_n-\Pi).
\end{equation}
This means that $V$
is a block-encoding of 
\begin{equation}
    X\otimes \Pi + I \otimes (I_n-\Pi)
\end{equation}
if $U_\Pi$ is a block-encoding of $\Pi$ with no error.
Let us define the  following quantum channel:
\begin{equation}
    \Lambda_\Pi(\rho)\coloneqq
    \sum_{i\in \{0,1\}^m}
    (I\otimes\bra{i}\otimes I_n)
    V
    (\ket{0^{m+1}}\bra{0^{m+1}}\otimes\rho)V^\dagger
    (I\otimes\ket{i}\otimes I_n).
\end{equation}
We also define 
\begin{equation}
    \Lambda_i(\rho)\coloneqq
    (I\otimes\bra{i}\otimes I_n)
    V
    (\ket{0^{m+1}}\bra{0^{m+1}}\otimes\rho)V^\dagger
    (I\otimes\ket{i}\otimes I_n).
\end{equation}
$\Lambda_0(\rho)$ is the quantum channel of \eqref{eq:BMmap}. 
Consider the case that what we can implement is not an exact block-encoding of $\Pi$ but an approximate one.
Let $\tilde{V}$ and $\tilde{\Lambda}_\Pi$ be the quantum circuit and channel which are similarly implemented as $V$ and $\Lambda_\Pi$ using $\tilde{U}_\Pi$ instead of $U_\Pi$.
The block-measurement theorem gives an upper bound on the diamond norm between $\tilde{\Lambda}_\Pi$ and ideal quantum channel $\Lambda_0$
:
\begin{theorem}[Block-measurement~\cite{rall2021faster}]
\label{them:blockmeasurement}
Let $\Pi$ be a projector, $V_\Pi=X\otimes\Pi+I\otimes (I_n-\Pi)$ be a unitary and $A$ be a hermitian matrix satisfying $\|\Pi-A^2\|\leq \epsilon$.
Also, $A$ has a block-encoding $\tilde{U}_\Pi$. 
Then the channel 
\begin{equation}
    \tilde{\Lambda}_\Pi(\rho)\coloneqq
    \sum_i
    (I\otimes\bra{i}\otimes I_n)
    \tilde{V}
    (\ket{0^{m+1}}\bra{0^{m+1}}\otimes\rho)\tilde{V}^\dagger
    (I\otimes\ket{i}\otimes I_n)
\end{equation}
approximates the quantum channel $\Lambda_0 (\rho)$ in diamond norm:
\begin{equation}
    \|\tilde{\Lambda}_\Pi-\Lambda_0\|_\diamond\leq 4\sqrt{2}\epsilon.
\end{equation}
\end{theorem}

\section{Quantum algorithm for additive error estimation of normalized Persistent Betti numbers}
\label{sec:main}
This section presents a quantum algorithm for estimating the persistent Betti numbers of arbitrary dimensions for a simplicial pair $K \hookrightarrow L$.  
We assume we have access to quantum membership oracles $O^K_q$ and $O^L_{q+1}$ that return whether a simplex is contained in $K$ and $L$ for any $\sigma\in\{0,1\}^n$ and $a\in \{0,1\}$, respectively, as
\begin{align}
    O^K_q|\sigma\rangle |a\rangle &=
    |\sigma\rangle |a\oplus f^K_q(\sigma)\rangle,\\
    O^L_{q+1}|\sigma\rangle |a\rangle &=
    |\sigma\rangle |a\oplus f^L_{q+1}(\sigma)\rangle,
\end{align}
where $f^K_q(\sigma)=1$ iff $\sigma\in \bar{S}^K_q$ and $f^L_{q+1}(\sigma)=1$ iff $\sigma\in \bar{S}^L_{q+1}$. 
For a simplicial complex with $n$ vertices, 
the number of all possible simplices is exponential in $n$. 
However, we can efficiently implement such a membership function for some constructions of the filtration that is commonly used in TDA, as we see in Appendix \ref{section:practical}. 
For example, 
the most straightforward way to implement the membership function for the Vietoris-Rips complex is to use the QRAM access to the $n\times n$ size adjacency matrix that represents the connectivity of the points.
Moreover, we can efficiently implement the membership function for the Vietoris-Rips complex by using a quantum circuit that returns the adjacency whenever we want because the adjacency matrix is of $n\times n$ size.

 Our main result is as follows:

\begin{theorem}
\label{them:main}
Let $K\hookrightarrow L$ be a simplicial pair that satisfies the following promises: \\\indent
(P1) $K$ is $q$-simplex dense: $d^K_q=n^K_q/{n \choose q+1}\in \Omega(1/poly(n))$\\\indent
(P2) $\Delta^L_{q,{\rm up}}(I^L_K,I^L_K)$ has inverse-polynomial gap: $\gamma_{\rm min}^q \in \Omega(1/poly(n))$\\\indent
(P3) the $q$-th persistent Laplacian $\Delta^{K,L}_q$ has inverse-polynomial gap: $\lambda_{\rm min}^q \in \Omega(1/poly(n))$.\\
Then,
given access to membership oracles $O^K_q$ and $O^L_{q+1}$, 
there is a quantum algorithm that outputs 1 with probability $\tilde{p}_1$ s.t. for any $\epsilon >0$ 
\begin{equation}
    \left|
    \frac{\beta^{K,L}_q}{n^K_q}-\tilde{p}_1
    \right|\leq \epsilon
\end{equation}
which uses $\mathcal{O}(n)+\mathcal{O}(\log{n})$ qubits, and $\tilde{\mathcal{O}}\left( poly(n)\left(\log{\frac{1}{\epsilon}}\right)^2 \right)$ number of $O^K_q,O^L_{q+1}$ and other quantum gates.
\end{theorem} 
As a consequence, we obtain the following corollary to estimate the persistent Betti numbers using the Chernoff-Heoffding inequality.

\begin{corollary}
For any simplicial pair $K\hookrightarrow L$ that satisfies the promises (P1) $\sim$ (P3) of Theorem \ref{them:main},
given access to membership oracles $O^K_q$ and $O^L_{q+1}$, 
we can efficiently estimate $\frac{\beta^{K,L}_q}{n^K_q}$ with additive error $\epsilon$ with confidence $\eta$ by taking 
$\mathcal{O}(\frac{1}{\epsilon^2}\ln(\frac{2}{\eta}))$ 
samples from the quantum algorithm in Theorem \ref{them:main}.
\end{corollary}

Without the promises (P1)$\sim$(P3), the complexity of our quantum algorithm can be stated as follows:
\begin{theorem}
\label{them:main2}
Let $K\hookrightarrow L$ be a simplicial pair.
For any $\epsilon_{sign}, \epsilon_{inv}, \epsilon_{rect}>0$ ,
given access to membership oracles $O^K_q$ and $O^L_{q+1}$, 
there is a quantum algorithm that outputs 1 with probability $\tilde{p}_1$ s.t. 
\begin{equation}
    \left|
    \frac{\beta^{K,L}_q}{n^K_q}-\tilde{p}_1
    \right|\leq  8\sqrt{2}\epsilon_\Pi + \epsilon_{sign},
\end{equation}
which uses $\mathcal{O}(n)+\mathcal{O}(\log{n})$ qubits and the following number of oracles and gates:
\begin{align}
\label{eq:oracle_gates}
    &O^K_q : \mathcal{O}\left(\sqrt{\frac{1}{d^K_q}}\log\left(\frac{1}{\epsilon_{sign}}\right)\right)
    +
    \mathcal{O}\left(
    \frac{q^4n^6}{(\gamma_{\rm min}^q)^2\lambda_{\rm min}^q}\log\left(\frac{1}{\epsilon_{rect}}\right)\log\left(\frac{1}{\gamma_{\rm min}^q \epsilon_{inv}}\right)\right),\\
    &O^L_{q+1}:\mathcal{O}\left(
    \frac{q^4n^6}{(\gamma_{\rm min}^q)^2\lambda_{\rm min}^q}\log\left(\frac{1}{\epsilon_{rect}}\right)\log\left(\frac{1}{\gamma_{\rm min}^q \epsilon_{inv}}\right)\right),\\
    &{\rm gates}: 
    \mathcal{O}\left(qn^2\sqrt{\frac{1}{d^K_q}}\log\left(\frac{1}{\epsilon_{sign}}\right)\right)
    +
    \tilde{\mathcal{O}}\left(
    \frac{q^4n^8}{(\gamma_{\rm min}^q)^2\lambda_{\rm min}^q}
    \log\left(\frac{1}{\epsilon_{rect}}\right)\log\left(\frac{1}{\gamma_{\rm min}^q \epsilon_{inv}}\right)\right).
\end{align}
Here, $\epsilon_\Pi=\epsilon_{rect} + \mathcal{O}\left(
{q^4n^4\sqrt{\epsilon_{inv}}\log(1/\epsilon_{rect})}
/({\lambda_{\rm min}^q\sqrt{\gamma_{\rm min}}})\right)$.
\end{theorem} 

{ In the following, we give the proof of Theorem~\ref{them:main} and Theorem~\ref{them:main2}. 
The proof consists of three parts. 
First, 
We propose the quantum algorithm in Section~\ref{sec:overview}, Section~\ref{subsec:statepreparation}, and Section~\ref{subsec:blockmeasurement}. 
(The construction of some unitaries are given in Section~\ref{section:statepreparation} and Section~\ref{section:projector}.)
Second, we give the error analysis in Section~\ref{sec:erroranalysis}. Finally, we give the complexity analysis in Section~\ref{sec:complexity}.}

\subsection{Description of the quantum algorithm}
\label{sec:overview}

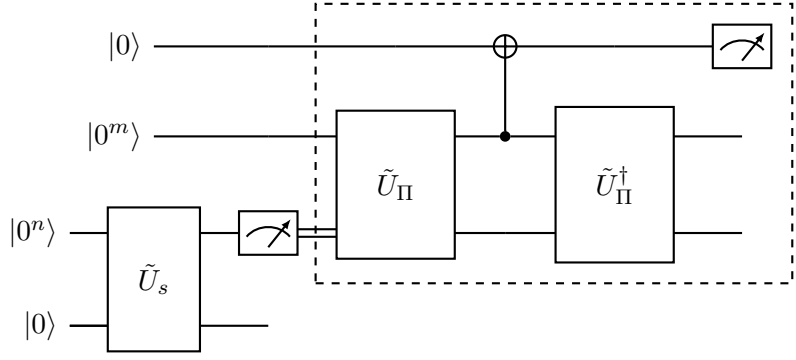
\begin{figure}[h]
\centering
\begin{quantikz}
&\lstick{\ket{0}}&\qw&\qw\gategroup[wires=3,steps=4,style={dashed}]{} & \targ{} &  \qw &\meter{} \\
&\lstick{\ket{0^m}}&\qw &\gate[2,cwires={2}]{\ \ \ \tilde{U}_{\Pi}\ \ \ } & \ctrl{} \vqw{-1} & \gate[wires=2]{\ \ \ \tilde{U}^\dagger_{\Pi}\ \ \ } &\qw \\
\lstick{\ket{0^n}}& \gate[wires=2]{\ \ \tilde{U}_s\ \ }& \meter{} &  & \qw & & \qw\\
\lstick{\ket{0}}&\qw&\qw&& &  & 
\end{quantikz}
\caption{Description of the quantum circuit for estimating the normalized persistent Betti numbers}
\label{fig:qc}
\end{figure}

The quantum circuit used in our algorithm is described in Fig.~\ref{fig:qc}.
In Fig.~\ref{fig:qc}, $\tilde{U}_s$ is a unitary for state preparation and $\tilde{U}_\Pi$ is a unitary that is a block-encoding of the projection onto the low-energy subspace of the persistent Laplacian. 
The $m$-qubits ($m\in \mathcal{O}(\log{n})$) are ancilla qubits for the block-encoding. 
The measurement procedure surrounded by the dashed box is the ``block-measurement'' procedure  \cite{rall2021faster} which we have reviewed in Section~\ref{subsec:review:blockmeasurement}.

\subsection{State Preparation $\tilde{U}_s$}
\label{subsec:statepreparation}
The state preparation unitary $\tilde{U}_s$ is the approximation of the following unitary $U_s$:
\[
U_s \ket{0^{n+1}} = |\phi^K_q\rangle\otimes \ket{1},
\]
where 
$|\phi^K_q\rangle \coloneqq \frac{1}{\sqrt{n^K_q}}\sum_{i\in [n^K_q]} |\sigma^K_q(i)\rangle$, which is the uniform superposition over the $q$-simplices of $K$. 
In Section~\ref{section:statepreparation}, we construct a unitary $\tilde{U}_s$ s.t. 
\begin{equation}
 \label{eq:statepreparationunitary}
    \left\||\phi^K_q\rangle\ket{1}- \tilde{U}_s|0^{n+1}\rangle\right\|
    \leq \epsilon_{sign}
\end{equation}
using 
$\mathcal{O}\left(qn^2\sqrt{1/d^K_q} \log{(1/\epsilon_{sign})}\right)$-number of gates and $\mathcal{O}\left(\sqrt{1/d^K_q}\log{(1/\epsilon_{sign})} \right)$-number of use of $O^K_q$.
Here, $d^K_q\coloneqq n^K_q/{n \choose q+1}$.
We denote $\tilde{U}_s\ket{0^{n+1}} := |\tilde{\psi}\rangle$.

We show that by measuring the state prepared by $\Tilde{U}_s$ in the computational basis as in Fig.~\ref{fig:qc}, we can approximately sample from the uniform mixture over the eigenvectors of $\Delta^{K,L}_q$. \footnote{
In the first version of the preprint of this paper~\cite{hayakawa2021quantumv1}, 
we copied the first register to another to obtain $\frac{1}{\sqrt{n^K_q}}\sum_{i\in [n^K_q]} |\sigma_q(i)\rangle\otimes\ket{1}\otimes |\sigma_q(i)\rangle$, and then traced out the latter registers to approximately obtain $\rho^K_q$. 
It is pointed out by an anonymous reviewer that this copy and discarding operation can be omitted by measuring in the computational basis and repeating the algorithm a number of times.
The copy and discarding operation is required if we use coherently estimate the amplitude using the amplitude estimation~\cite{brassard2002quantum} for example.
}
By measuring the first $n$-qubit of $U_s \ket{0^{n+1}}$, the state will collapse to some $|\sigma^K_q(i)\rangle$ approximately uniform randomly. 
Repeating the quantum circuit of Fig.~\ref{fig:qc} and taking samples multiple times from the collapsed state is equivalent to sampling from a state described by a density matrix $\tilde{\rho}^K_q$ which is close to $\rho^K_q\coloneqq \frac{1}{n^K_q}\sum_{i\in [n^K_q]} |\sigma_q(i)\rangle\langle \sigma_q(i)|$.
This $\rho^K_q$ is also the uniform mixture of the eigenvectors of $\Delta^{K,L}_q$, i.e.,
$\rho^K_q=\frac{1}{n^K_q}\sum_{i\in [n^K_q]} |\psi_i\rangle\langle \psi_i|$, where we denote the eigenvectors of $\Delta^{K,L}_q$ as  $|\psi_1\rangle,...,|\psi_{n^K_q}\rangle$.
This is because $\forall i$, $|\sigma_q(i)\rangle$ 
can be written as $|\sigma_q(i)\rangle=\sum_{j} U_{i,j}\ket{\psi_j}$ for some unknown unitary $U$ and
\begin{equation}
    \rho^K_q=\frac{1}{n^K_q}\sum_{i}|\sigma_q(i)\rangle\langle \sigma_q(i)| 
=\frac{1}{n^K_q}\sum_{j,j'} \left(\sum_i U_{i,j}U^*_{i,j'}\right)
|\psi_{j}\rangle\langle \psi_{j'}| 
=\frac{1}{n^K_q}\sum_{j} |\psi_j\rangle\langle \psi_j|.
\end{equation}
We give an upper bound on 
$\|{\rho}^K_q-\tilde{\rho}^K_q\|_1$, which we use in the error analysis later.
Using purification, $\tilde{\rho}^K_q$ can be written as 
\[
\tilde{\rho}^K_q = {\rm Tr}_s \left[ 
U_{\rm copy} \left(|\tilde{\psi}
\rangle\langle \tilde{\psi}|\otimes\ket{0^n}\bra{0^n}\right)U_{\rm copy}^\dagger
\right],
\]
where $U_{\rm copy}$ is a unitary that  copies the first $n$-qubit register to the last $n$-qubit register, and the subscript $s$ means the last $n$-qubit register. 
${\rho}^K_q$ can be similarly written as 
\[
{\rho}^K_q = {\rm Tr}_s \left[ 
U_{\rm copy} \left(|{\psi}
\rangle\langle {\psi}|\otimes\ket{0^n}\bra{0^n}\right)U_{\rm copy}^\dagger
\right],
\]
where $\ket{\psi}\coloneqq \ket{\phi^K_q}\otimes\ket{1}$. 
Then
\begin{align}
    \label{eq:error:statepreparation}
    \left\|{\rho}^K_q-\tilde{\rho}^K_q\right\|_1
    &=
\left\|
    {\rm Tr}_s 
    \left[
        U_{\rm copy} (|{\psi}
        \rangle\langle {\psi}|\otimes\ket{0^n}\bra{0^n}
        -
        |\tilde{\psi}
        \rangle\langle \tilde{\psi}|\otimes\ket{0^n}\bra{0^n}
        )U_{\rm copy}^\dagger
    \right]
\right\|_1\\ 
&\leq
\left\|
        U_{\rm copy} (|{\psi}
        \rangle\langle {\psi}|\otimes\ket{0^n}\bra{0^n}
        -
        |\tilde{\psi}
        \rangle\langle \tilde{\psi}|\otimes\ket{0^n}\bra{0^n}
        )U_{\rm copy}^\dagger
\right\|_1\\
&= 
\left\|
|{\psi}
\rangle\langle {\psi}|\otimes\ket{0^n}\bra{0^n}
-
|\tilde{\psi}
\rangle\langle \tilde{\psi}|\otimes\ket{0^n}\bra{0^n}
\right\|_1\\
&= 
\left\|
|{\psi}
\rangle\langle {\psi}|
-
|\tilde{\psi}
\rangle\langle \tilde{\psi}|
\right\|_1 
=\sqrt{1-|\bra{\psi}\tilde{\psi}\rangle|^2}
\leq  \epsilon_{sign}.
\end{align}

\subsection{Block-measurement with respect to $\tilde{U}_\Pi$}
\label{subsec:blockmeasurement}

{
In Section~\ref{section:projector}, we provide the implementation of $\tilde{U}_\Pi$ which is a $(1,m,\epsilon_\Pi)$-block-encoding of  $\Pi\coloneqq \sum_{j:\lambda_j=0}
\ket{\psi_j}\bra{\psi_j}$ assuming that the smallest non-zero eigenvalue of $\Delta^{K,L}_q$ is $\lambda_{\rm min}^q$.
We also assume that the smallest non-zero eigenvalue of the submatrix $\Delta^L_{q,{\rm up}}(I^L_K,I^L_K)$ is $\gamma_{\rm{min}}^q$.
The number of oracles and gates that are used in the construction of $\tilde{U}_\Pi$ is as follows:
\begin{align}
    &O^K_q : 
    \mathcal{O}\left(
    \frac{q^4n^6}{(\gamma_q)^2\lambda_q}\log\left(\frac{1}{\epsilon_{rect}}\right)\log\left(\frac{1}{\gamma_q \epsilon_{inv}}\right)\right),\\
    &O^L_{q+1}:\mathcal{O}\left(
    \frac{q^4n^6}{(\gamma_q)^2\lambda_q}\log\left(\frac{1}{\epsilon_{rect}}\right)\log\left(\frac{1}{\gamma_q \epsilon_{inv}}\right)\right),\\
    &{\rm gates}: 
    \tilde{\mathcal{O}}\left(
    \frac{q^4n^8}{(\gamma_q)^2\lambda_q}
    \log\left(\frac{1}{\epsilon_{rect}}\right)\log\left(\frac{1}{\gamma_q \epsilon_{inv}}\right)\right).
\end{align}
Here, $\lambda_q$ and $\gamma_q$ are real numbers s.t. $0<\lambda_q<\lambda_{\rm min}^q$ and 
$0<\gamma_q<\gamma_{\rm{min}}^q$, and 
$\epsilon_{rect}, \epsilon_{inv}>0$ are real numbers s.t. 
$\epsilon_\Pi=\epsilon_{rect} + \mathcal{O}\left(
    {q^4n^4\sqrt{\epsilon_{inv}}\log(1/\epsilon_{rect})}
    /({\lambda_q\sqrt{\gamma_q}})\right)$.}

$U_\Pi$ satisfies 
\begin{equation}
      \left\|\Pi-
      \left((\bra{0^m}\otimes I_{n+1}) U_\Pi ( \ket{0^m}\otimes I_{n+1})
      \right)^2
      \right\|
    \leq
    2\epsilon_\Pi-\epsilon_\Pi^2\leq 2\epsilon_\Pi.
\end{equation}
We perform block-measurement with this $\tilde{U}_\Pi$. 
Let $\Lambda_\Pi$ be a quantum channel that maps as 
\begin{equation}
    |0\rangle\otimes|\psi\rangle
    \rightarrow
    |1\rangle\otimes\Pi |\psi\rangle 
    +|0\rangle\otimes (I-\Pi)|\psi\rangle.
\end{equation}
This is a quantum channel realized by ideal block-encoding $U_\Pi$ of $\Pi$ with no error, which we denote $\Lambda_\Pi$.
From Theorem~\ref{them:blockmeasurement}, 
the quantum channel realized by the block-measurement with $\tilde{U}_\Pi$ satisfies
\begin{equation}
\label{eq:error:BM}
    \|\tilde{\Lambda}_\Pi-\Lambda_\Pi\|_\diamond\leq 8\sqrt{2}
    \epsilon_\Pi.
\end{equation} 
We use this inequality in error analysis in the next subsection.

\subsection{Error analysis}
\label{sec:erroranalysis}

It can be seen that for the ideal state preparation and the ideal block-measurement, the probability of outputting $1$ is
\begin{align}
    p_1\coloneqq &
    {\rm Tr}_{AB}
    \left[
    (|1\rangle\langle 1|\otimes I_{m+n})
    V
    \Big( \ket{0^{m+1}}\bra{0^{m+1}}\otimes \rho^K_q \Big) 
    V^\dagger
    \right]\\
    =&
    {\rm Tr}_{A}
    \left[
    (|1\rangle\langle 1|\otimes I_{n})
    \Lambda_0
    \Big( \ket{0}\bra{0}\otimes \rho^K_q \Big) 
    \right]\\
    =&
    \frac{|\{i:\lambda_i=0\}|}{n^K_q}=\frac{\rm{nullity}(\Delta^{K,L}_q)}{n^K_q},
\end{align}
where we have denoted the $m$-qubit register as $A$ and other registers as $B$, and $V$ is the block-measurement circuit of Fig.~\ref{fig:blockmeasurement} with $U_\Pi$. 
The actual probability that the measurement outcome is $1$ in the final measurement of Fig.~\ref{fig:qc} is
\begin{align}
    \tilde{p}_1
=&
    {\rm Tr}_{AB}
    \left[
    (|1\rangle\langle 1|\otimes I_{m+n})
    \tilde{V}
    \Big( \ket{0^{m+1}}\bra{0^{m+1}}\otimes \tilde{\rho}^K_q \Big) 
    \tilde{V}^\dagger
    \right]\\
    =&
    {\rm Tr}_{A}
    \left[
    (|1\rangle\langle 1|\otimes I_{n})
    \tilde{\Lambda}_{\Pi}
    \Big( |0\rangle\langle 0|\otimes \tilde{\rho}^K_q \Big)
    \right],
\end{align}
where $\tilde{V}$ is the block-measurement circuit of Fig.~\ref{fig:blockmeasurement} with $\tilde{U}_\Pi$. 
Let $\mathcal{M}=|1\rangle\langle1|\otimes I_n$. 
The difference between the ideal probability and the actual probability can be calculated as follows:
\begin{align}
    |p_1-&\tilde{p}_1|
    =
    %1
    \left|
    {\rm Tr}
    \left[\mathcal{M}\Lambda_0(|0\rangle\langle 0|\otimes\rho^K_q)
    \right]
    -
    {\rm Tr}
    \left[\mathcal{M}\tilde{\Lambda}_{\Pi}(|0\rangle\langle 0|\otimes\tilde{\rho}^K_q)
    \right]\right|\\
    %2
    &\leq
    \left|
    {\rm Tr}
    \left[\mathcal{M}
    \left(\Lambda_0(|0\rangle\langle 0|\otimes\rho^K_q)
    -
    \tilde{\Lambda}_{\Pi}(|0\rangle\langle 0|\otimes\rho^K_q)
    \right)
    \right]\right|
    +
    \left|
    {\rm Tr}
    \left[\mathcal{M}\tilde{\Lambda}_{\Pi}\left(|0\rangle\langle 0|\otimes(\rho^K_q-\tilde{\rho}^K_q\right)\right]
    \right|\\
    %3
    &\leq
    \left\|
    (\Lambda_0-\tilde{\Lambda}_{\Pi})\left(|0\rangle\langle 0|\otimes\rho^K_q\right)\right\|_1
    +
    \left\|
    \rho^K_q-\tilde{\rho}^K_q
    \right\|_1\\
    &\leq
    \|\Lambda_0-\tilde{\Lambda}_\Pi\|_\diamond
    +
    \left\|
    \rho^K_q-\tilde{\rho}^K_q
    \right\|_1\\
    &\leq
    8\sqrt{2}\epsilon_\Pi + {\epsilon_{sign}},
\end{align}
where we have used 
\eqref{eq:error:statepreparation}
and \eqref{eq:error:BM}.

\subsection{Analysis of the number of gates and the use of the oracles.}
\label{sec:complexity}

The total number of gates and use of oracles can be summarized as follows:
\begin{align}
    &O^K_q : \mathcal{O}\left(\sqrt{\frac{1}{d^K_q}}\log\left(\frac{1}{\epsilon_{sign}}\right)\right)
    +
    \mathcal{O}\left(
    \frac{q^4n^6}{(\gamma_q)^2\lambda_q}\log\left(\frac{1}{\epsilon_{rect}}\right)\log\left(\frac{1}{\gamma_q \epsilon_{inv}}\right)\right),\\
    &O^L_{q+1}:\mathcal{O}\left(
    \frac{q^4n^6}{(\gamma_q)^2\lambda_q}\log\left(\frac{1}{\epsilon_{rect}}\right)\log\left(\frac{1}{\gamma_q \epsilon_{inv}}\right)\right),\\
    &{\rm gates}: 
    \mathcal{O}\left(qn^2\sqrt{\frac{1}{d^K_q}}\log\left(\frac{1}{\epsilon_{sign}}\right)\right)
    +
    \tilde{\mathcal{O}}\left(
    \frac{q^4n^8}{(\gamma_q)^2\lambda_q}
    \log\left(\frac{1}{\epsilon_{rect}}\right)\log\left(\frac{1}{\gamma_q \epsilon_{inv}}\right)\right).
\end{align}
If we want to make $|p_1-\tilde{p}_1|\leq \epsilon$, we can take $\epsilon_\Pi\in \mathcal{O}(\epsilon)$, {$\epsilon_{sign}\in\mathcal{O}(\epsilon)$}, $\epsilon_{rect}\in \mathcal{O}(\epsilon)$ and \[\epsilon_{inv}\in \mathcal{O}\left(\frac{(\lambda_q)^2\gamma_q\epsilon^2}{q^8n^8\log{(1/\epsilon)}}\right).\]
Therefore, if $d^K_q=n^K_q/{n \choose q+1}\in \Omega(1/poly(n))$, $\gamma_{\rm min}^q \in \Omega(1/poly(n))$ and $\lambda_{\rm min}^q \in \Omega(1/poly(n))$,
the quantum circuit uses $\tilde{\mathcal{O}}\left( poly(n)\left(\log{\frac{1}{\epsilon}}\right)^2 \right)$ number of $O^K_q,O^L_{q+1}$ and other quantum gates.

\section{Implementation of the state preparation unitary}
\label{section:statepreparation}

{
In this section, we provide the implementation of the state preparation unitary $\tilde{U}_s$ of eq.~\eqref{eq:statepreparationunitary}.}
In order to implement $\tilde{U}_s$,
we use a unitary $P_q$ s.t.
\begin{equation}
\label{eq:dicke}
    P_q \ket{0^n} = \frac{1}{\sqrt{n \choose q+1}}\sum_{x\in W_q} |x\rangle,
\end{equation}
where $W_q$ is the set of Hamming weight $q+1$ $n$-bit strings. 
In \cite{gunn2019review}, 
the combinatorial number system~\cite{siddique2016proof} is used to implement such a unitary. 
The uniform superposition of the Hamming weight $k$ states is also known as the Dicke state \cite{dicke1954coherence}. 
Instead of following the construction of \cite{gunn2019review} using the combinatorial number system, 
we can use the construction of
\cite{bartschi2019deterministic}
for the Dicke state. 
Then, we can implement $P_q$ with $\mathcal{O}(qn)$ gates and $\mathcal{O}(n)$ depth without using any ancilla qubits. 
(The gate complexity of \cite{gunn2019review} is similar, but it additionally requires $\tilde{\mathcal{O}}(qn^2)$ time of computation for preparing a look-up table.)

Then by adding a qubit and applying the membership oracle $O^K_q$ to the quantum state of \eqref{eq:dicke}, we get
\begin{equation}
    \frac{1}{\sqrt{n \choose q+1}}
    \left(\sum_{x\in S^K_q} |x\rangle|1\rangle +
    \sum_{x\notin S^K_q} |x\rangle|0\rangle\right)
    =
     \sqrt{d^K_q}|\phi^K_q\rangle\ket{1} +
    \frac{1}{\sqrt{n \choose q+1}}\sum_{x\notin S^K_q} |x\rangle|0\rangle
    .
\end{equation}
We call $d^K_q$ the $q$-simplex density of $K$. 
Let $\tilde{\Pi}=I_n \otimes |1\rangle\langle 1|$, $\Pi=\ket{0^{n+1}}\bra{0^{n+1}}$ and $U=O^K_q P_q$. 
Then $\tilde{\Pi}U\Pi=\sqrt{d^K_q}(|\phi^K_q\rangle\ket{1})\langle 0^{n+1}|$.
Now, we can use Theorem~\ref{theorem:fixedpointamp} to implement a unitary $\tilde{U}_s$ s.t. 
\begin{equation}
    \left\||\phi^K_q\rangle\ket{1}- \tilde{U}_s|0^{n+1}\rangle\right\|
    \leq \epsilon_{sign}
\end{equation}
with $\mathcal{O}
\left
(
\sqrt{1/d^K_q}\log{(1/\epsilon_{sign})} \right)$ use of $U$,$U^\dagger$, $\rm{C}_{\tilde{\Pi}}\rm{NOT}$, $\rm{C}_{{\Pi}}\rm{NOT}$, and single qubit gates. 
Each of the $\rm{C}_{{\Pi}}\rm{NOT}$ gates, which are the generalized Toffoli gates, can be decomposed into $\mathcal{O}(n)$ number of elementary gates with an ancilla qubit \cite{he2017decompositions}. 
Therefore, $\tilde{U}_s$ can be implemented using 
$\mathcal{O}\left(qn^2\sqrt{1/d^K_q} \log{(1/\epsilon_{sign})}\right)$-number of gates and $\mathcal{O}\left(\sqrt{1/d^K_q}\log{(1/\epsilon_{sign})} \right)$-number of use of $O^K_q$.

\section{Implementation of the block-encoding of the projector $\tilde{U}_\Pi$}
\label{section:projector}

{
In this section, we give a construction of $\tilde{U}_\Pi$ which is a $(1,m,\epsilon_\Pi)$-block-encoding of  $\Pi$.
We first show the implementation of the block-encoding of the persistent Laplacian
\begin{equation}
 \Delta_q^{K,L}=
 \underbrace{
 \partial_{q+1}^{L,K}\circ (\partial_{q+1}^{L,K})^{*}
 }_{\Delta^{K,L}_{q,{\rm up}}}
 +
 \underbrace{
 (\partial_{q}^K)^{*}\circ \partial_{q}^K 
 }_{\Delta^{K}_{q,{\rm down}}}
\end{equation}
in Section~\ref{imp:1}.
Then we show the construction of a $(1,m,\epsilon_\Pi)$-block-encoding of  $\Pi$ in Section~\ref{imp:4} using the block-encoding of $\Delta_q^{K,L}$. }

The implementation of $\tilde{U}_\Pi$ includes the inversion of matrices. 
More precisely, our algorithm includes the implementation of the Schur complement of a matrix that contains the implementation of the Moore-Penrose pseudo-inverse. 
The implementation of the Schur complement may be of independent interest. 
The fact that our algorithm includes the matrix inversion implies that it shares the characteristics of both the HHL algorithm \cite{harrow2009quantum} and the LGZ algorithm.

\subsection{Implementation of the block-encoding of $\Delta^K_q$}
\label{imp:1}
We provide the procedure of implementing the block-encoding of $\Delta^K_q$ in the following.

\subsubsection{
Block-encoding of $\Delta^K_{q,\rm{down}}=(\partial_{q}^K)^{*}\circ \partial_{q}^K$}
{Let $\tilde{n}\coloneqq 2^{\lceil \log{n}\rceil}$, and $\tilde{q}\coloneqq 2^{\lceil \log{(q+1)}\rceil}$.
In Appendix~\ref{sec:appendix:boundaryopeartor}}, we construct a unitary $U^K_q$ which is an $({\tilde{n}\tilde{q}},a,0)$-block-encoding of $\partial^K_q$. 
$U^K_q$ is implemented using $\mathcal{\tilde{O}}(n^2)$-number of gates and $\mathcal{O}(1)$-use of $O^K_q$. Here $a=\mathcal{O}(\log (n))$.

It follows that 
$(U^K_q)^\dagger$ is an $({\tilde{n}\tilde{q}},a,0)$-block-encoding of $(\partial^K_q)^*$ because the matrix representation of $(\partial^K_q)^*$ is the transpose of the matrix representation of $\partial^K_q$ whose elements are real. 
We can implement the block-encoding of $\Delta^K_{q,\rm{down}}$ 
using the following lemma:

\begin{lemma}[Product of block-encoded matrices (Lemma 53 of~\cite{gilyen2019quantum})]
\label{lemma:product}
If $U$ is an $(\alpha,a,\delta)$-block-encoding of an $s$-qubit operator $A$,
and $V$ is a $(\beta,b,\epsilon)$-block-encoding of a
$s$-qubit operator $B$ then $(I_b\otimes U)(I_a\otimes V)$ is an $(\alpha\beta,a+b,\alpha\epsilon+\beta\delta)$-block-encoding of $AB$. 
\end{lemma}
Using this lemma, we can implement a unitary
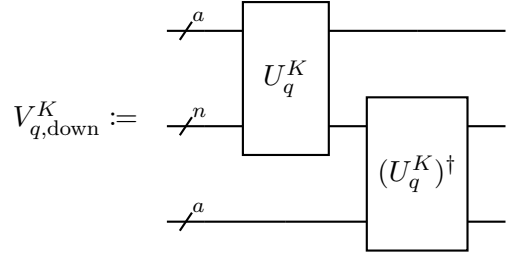
\begin{center}
$V^K_{q,\rm{down}}\coloneqq$
\begin{quantikz}[transparent]
&\qwbundle
{a}  & \gate[2]{\ U^K_q\ } & \qw & \qw \\
&\qwbundle
{n}  &  &\gate[2]{(U^K_q)^\dagger} & \qw \\
& \qwbundle
{a}  &\qw&& \qw
\end{quantikz}
\end{center}
which is an $({\tilde{n}^2\tilde{q}^2},2a,0)$-block-encoding of $\Delta^K_{q,\rm{down}}$. 
From the construction above, $V^K_{q,\rm{down}}$ can be implemented with $\mathcal{\tilde{O}}(n^2)$ gates and $\mathcal{O}(1)$ use of $O^K_q$.

\subsubsection{Implementation of the submatrices of $\Delta^L_{q,\rm{up}}$ and its Schur complement }
\label{imp:2}

Let us define the submatrices
\begin{align}
\label{eq:delta1to4}
    &\Delta_1\coloneqq \Delta^L_{q,{\rm up}}([n^K_q],[n^K_q])\\
    &\Delta_2\coloneqq \Delta^L_{q,{\rm up}}([n^K_q],I^L_K)\\
    &\Delta_3\coloneqq \Delta^L_{q,{\rm up}}(I^L_K,[n^K_q])\\
    &\Delta_4\coloneqq \Delta^L_{q,{\rm up}}(I^L_K,I^L_K)
\end{align}
for convenience.
By Theorem~\ref{theorem:schur}, 
\begin{equation}
    \Delta^{K,L}_{q,{\rm up}}=\Delta^L_{q,{\rm up}}/\Delta^L_{q,{\rm up}}(I^L_K,I^L_K)
    =\Delta_1-\Delta_2\Delta_4^+\Delta_3.
\end{equation}
{Let us also define ${q}'\coloneqq 2^{\lceil \log{(q+2)} \rceil}$.}
We can similarly implement a unitary $V^L_{q,\rm{up}}$ which is a  $(\tilde{n}^2{q'}^2,2b,0)$-block-encoding of $\Delta^L_{q,\rm{up}}:\mathcal{H}^L_q\rightarrow \mathcal{H}^L_q$ where $b=\mathcal{O}(\log (n))$ with using $\mathcal{\tilde{O}}(n^2)$ number of gates and $\mathcal{O}(1)$ number of $O^L_{q+1}$.

In order to implement the Schur complement $\Delta^L_{q,{\rm up}}/\Delta^L_{q,{\rm up}}(I^L_K,I^L_K)$, we need to access the submatrices of $\Delta^L_{q,\rm{up}}$. 
The membership oracle $O^K_q:\ket{x}\ket{0}\rightarrow \ket{x}\ket{f^K_q(x)}$ can be used to access those submatrices. 
Let $\tilde{O}^K_q:\ket{x}\ket{0}\rightarrow \ket{x}\ket{f^K_q(x)\oplus 1}$.
Then, let us introduce
\begin{align}
    &V_1\coloneqq (I\otimes\tilde{O}^K_q \otimes I_{2b}) (I_2\otimes V^L_q) (\tilde{O}^K_q  \otimes I\otimes I_{2b}),\\
    &V_2\coloneqq (I\otimes\tilde{O}^K_q \otimes I_{2b}) (I_2\otimes V^L_q) (O^K_q  \otimes  I\otimes I_{2b}),\\
    &V_3\coloneqq (I\otimes O^K_q \otimes I_{2b}) (I_2\otimes V^L_q) (\tilde{O}^K_q \otimes I\otimes I_{2b}),\\
    &V_4\coloneqq (I\otimes O^K_q \otimes I_{2b}) (I_2\otimes V^L_q) (O^K_q \otimes I\otimes I_{2b}).
\end{align}
By applying $\tilde{O}^K_q$ and postselecting the ancilla register to $\ket{0}$, we can restrict the space to that is spanned by $\{|\sigma^K_q(i)\rangle\}_{i\in [n^K_q]}$. 
Similarly, by applying ${O}^K_q$ with postselecting the ancilla register to $\ket{0}$, we can restrict the space to that is not spanned by $\{|\sigma^K_q(i)\rangle\}_{i\in [n^K_q]}$. 
Therefore, $V_1$, $V_2$, $V_3$ and $V_4$ 
are $({\tilde{n}^2{q'}^2},2b+2,0)$-block-encodings of $\Delta_1$, $\Delta_2$, $\Delta_3$ and $\Delta_4$, respectively. 
$V_1 \sim V_4$ can be implemented using $\mathcal{\tilde{O}}(n^2
)$-gates and $\mathcal{O}(1)$-use of $O^K_q$ and $O^L_{q+1}$.

Let $\gamma_{\rm{min}}^q$ be the smallest non-zero eigenvalue of $\Delta_4$ and let $\kappa =\frac{{\tilde{n}^2{q'}^2}}{\gamma_q}$ for $\gamma_q$ s.t. $0<\gamma_q<\gamma_{\rm{min}}^q$.
Then, using Theorem \ref{theorem:pseudoinverse},
we can  implement a 
$(\frac{2}{\gamma_q},2b+3,\epsilon_{inv})$-
block-encoding of $\Delta_4^+$
with $\mathcal{O}(\frac{q^2n^2}{\gamma_q}\log(\frac{1}{\gamma_q\epsilon_{inv}}))$ use of $V_4$, $V_4^\dagger$ and $\mathcal{\tilde{O}}(\frac{q^2n^2}{\gamma_q}\log(\frac{1}{\gamma_q\epsilon_{inv}}))$ other elementary gates.

We can implement a unitary $V_+$ which is a $(\frac{2{\tilde{n}^4{{q}'}^4}}{\gamma_q},6b+7,{\tilde{n}^4{{q}'}^4}\epsilon_{inv})$-block-encoding of $\Delta_2\Delta_4^+\Delta_3$ using Lemma~\ref{lemma:product}.

\begin{remark}
As in Theorem~\ref{theorem:schur}, 
$\Delta^{K,L}_{q,{\rm up}} = \Delta^{L}_{q,{\rm up}}/\Delta^{L}_{q,{\rm up}}(I^L_K,I^L_K)$
,where $I^L_K\coloneqq [n^L_q]\backslash [n^K_q]$. 
However, we do not need to know the values of $n^k_q$ and $n^L_q$ to run our algorithm because we can implement the block-encoding of submatrices by $V_1$, $V_2$, $V_3$ and $V_4$
{
using the membership oracles.
The membership oracles have been used to effectively implement the restrictions to the corresponding subspaces of $I_k^L$ coherently, bypassing the need to classically learn the values of $I_k^L$.}
\end{remark}

\subsubsection{Linear combination of block-encoding unitaries}
Finally, we linearly combine the block-encoding (BE) of $\Delta_1$, $\Delta_2\Delta_4^+\Delta_3$ and $\Delta^K_{q,\rm{down}}$. 
We implement such a linear combination in a way similar to Lemma 52 of \cite{gilyen2019quantum}.
What we have prepared so far are the following unitaries:
\begin{equation}
\begin{array}{llcl}
U_0\coloneqq V_1 &\Leftrightarrow &(\alpha_0\coloneqq {\tilde{n}^2{{q}'}^2},2b+2,0)-\rm{BE}\ \rm{of}\ &\Delta_1\\
U_1\coloneqq V_+ &\Leftrightarrow &(\alpha_1\coloneqq \frac{2{\tilde{n}^4{{q}'}^4}}{\gamma_q},6b+7,{\tilde{n}^4{{q}'}^4}\epsilon_{inv})-\rm{BE}\ \rm{of}\ &
\Delta_2\Delta_4^+\Delta_3\\
U_2\coloneqq V^K_q  &\Leftrightarrow &(\alpha_2\coloneqq {\tilde{n}^2\tilde{q}^2},2a,0)-\rm{BE}\ \rm{of}\ &\Delta^K_{q,\rm{down}} ,
\end{array}
\end{equation}
where we have introduced $U_0,U_1,U_2$ and $\alpha_0,\alpha_1,\alpha_2$ for the convenience of notation.
What we implement is a block-encoding of 
\begin{equation}
    \Delta^{K,L}_q= \Delta_1-\Delta_2\Delta_4^+\Delta_3+\Delta^K_{q,\rm{down}}.
\end{equation}
Let  $\beta = \alpha_0+\alpha_1+\alpha_2$ and $P_R, P_L$ be state preparation unitaries such that 
\begin{align}
    &P_R\ket{0}\rightarrow \frac{1}{\sqrt{\beta}}(\sqrt{\alpha_0}\ket{0}+\sqrt{\alpha_1}\ket{1}+\sqrt{\alpha_2}\ket{2}),\\
    &P_L\ket{0}\rightarrow \frac{1}{\sqrt{\beta}}(\sqrt{\alpha_0}\ket{0}-\sqrt{\alpha_1}\ket{1}+\sqrt{\alpha_2}\ket{2}).
\end{align}
Suppose $W$ is a unitary such that for all $j\in{0,1,2}$ and any state $\ket{\psi}$,
\begin{equation}
    W:\ket{j}\ket{\psi}
    \rightarrow 
    \ket{j} U_j\ket{\psi}.
\end{equation}
Let $\tilde{W}=(P_L^\dagger\otimes I_{6b+7}\otimes I_n)W(P_R\otimes I_{6b+7}\otimes I_n)$.
Then, this satisfies
\begin{equation}
    \left\|\left(\langle0^{6b+9}|\otimes I_n\right) \tilde{W} \left(|0^{6b+9}\rangle \otimes I_n\right)
    - \frac{\Delta^{K,L}_q}{\beta}
    \right\|\leq 
    \frac{{\tilde{n}^4{{q}'}^4} \epsilon_{inv}}{\beta}
\end{equation}
and $\tilde{W}$ is a
$(\beta,6b+9,{\tilde{n}^4{{q}'}^4}\epsilon_{inv})$-block-encoding of $\Delta^{K,L}_q$, and $\beta \in \mathcal{O}(\frac{q^4n^4}{\gamma_q})$. 
Let us denote 
\begin{equation}
    \left(\langle0^{6b+9}|\otimes I_n\right) \tilde{W} \left(|0^{6b+9}\rangle \otimes I_n\right) = \tilde{\Delta}^{K,L}_q /\beta
\end{equation}
for $\tilde{\Delta}^{K,L}_q$ s.t. $\left\|\Delta^{K,L}_q-\tilde{\Delta}^{K,L}_q\right\|\leq {\tilde{n}^4{{q}'}^4}\epsilon_{inv}$.

\subsection{Implementing the block-encoding of the projector}
\label{imp:4}
Using the block-encoding of $\Delta^{K,L}_q$, we implement the block-encoding of the projector onto the zero energy space of $\Delta^{K,L}_q$: $\Pi= \sum_{j:\lambda_j=0}
\ket{\psi_j}\bra{\psi_j}$. 
Let $\lambda_{\rm min}^q$ be the smallest non-zero eigenvalue of $\Delta^{K,L}_q$. 
In \cite{gilyen2019quantum}, the following polynomial approximation of the rectangle function is shown:
\begin{lemma}[Lemma 29, \cite{gilyen2019quantum}]
\label{lemma:rectangle}
Let $\delta,\epsilon\in(0,1/2)$ and $t\in [-1,1]$. There exist an even polynomial $P\in {\mathbb R}[x]$ of degree ${\mathcal O}(\log{(1/\epsilon)}/\delta)$, such that $|P(x)|\leq 1$ for all $x\in[-1,1]$, and 
\begin{equation}
\left\{
    \begin{array}{ll}
    P(x)\in [0,\epsilon], & \forall x\in[-1,-t-\delta]\cup [t+\delta,1],\ and\\
    P(x)\in [1-\epsilon,1], & \forall x\in [-t+\delta,t-\delta].
    \end{array}
\right.
\end{equation}
\end{lemma}
We take $t=\frac{\lambda_{\rm min}^q}{2\beta}$ and $\delta=\frac{\lambda_q}{2\beta}$ for such $\lambda_q$ that satisfies $0<\lambda_q<\lambda_{\rm min}^q$. 
Because the block-encoding of $\Delta^{K,L}_q$ is not exact, we need to introduce the following lemma on the robustness of singular value transformation:

\begin{lemma}[Lemma 22 of \cite{gilyen2019quantum}]
\label{lemma:robustness}
If $P\in\mathbb{C}[x]$ is a degree-$d$ polynomial satisfying the following conditions:
\begin{itemize}
    \item $P$ has parity-($d$ mod 2),
    \item $\forall x \in [-1,1] : |P(x)|\leq 1$,
    \item $\forall x \in (-\infty,-1]\cup [1,\infty):|P(x)|\geq 1$,
    \item if $d$ is even, then $\forall x\in {\mathbb R}:P(ix)P^*(ix)\geq 1$
\end{itemize}
and moreover, 
$A,\tilde{A}\in \mathbb{C}^{\tilde{n}\times n}$ are matrices of operator norm at most 1, then we have that 
\begin{equation}
\left\| 
P^{\rm SV}(A)-P^{\rm SV}(\tilde{A})
\right\|
\leq
4d\sqrt{\|A-\tilde{A}\|}.
\end{equation}
\end{lemma}
Using Lemma~\ref{lemma:rectangle} and  Corollary~\ref{corol:qsvtreal},
we can implement a unitary $\tilde{U}_\Pi$ 
which is a QSVT of $\tilde{\Delta}^{K,L}_q/\beta$ with respect to the rectangle function: 
\begin{equation}
    P^{\rm(SV)}\left(\tilde{\Delta}^{K,L}_q/\beta\right) = 
    \left(\langle0^{6b+10}|\otimes I_n\right) \tilde{U}_\Pi \left(|0^{6b+10}\rangle \otimes I_n\right).
\end{equation}
Using Lemma~\ref{lemma:robustness} it can be seen that
\begin{equation}
    \left\| P^{\rm(SV)}\left({\Delta}^{K,L}_q/\beta\right) -P^{\rm(SV)}\left(\tilde{\Delta}^{K,L}_q/\beta\right) \right\|
    \leq 
    4d \sqrt{\|{\Delta}^{K,L}_q/\beta-\tilde{\Delta}^{K,L}_q/\beta\|},
\end{equation}
where $d\in \mathcal{O}(\beta\log(1/\epsilon_{rect})/\lambda_q)$.
$P^{\rm(SV)}\left({\Delta}^{K,L}_q/\beta\right)$ is $\epsilon_{rect}$-close to $\Pi$ in spectral norm.
Therefore, 
$\tilde{U}_\Pi$ is a $(1,m\coloneqq6b+10,\epsilon_\Pi\coloneqq\epsilon_{rect} + 4d\sqrt{{\tilde{n}^4{{q}'}^4}\epsilon_{inv} /\beta})$-block-encoding of $\Pi$.
It can be seen that
\begin{equation}
    4d\sqrt{{\tilde{n}^4{{q}'}^4}\epsilon_{inv}/\beta}
    =
    \mathcal{O}(n^2q^2\sqrt{\beta\epsilon_{inv}}\log(1/\epsilon_{rect})/\lambda_q)
    =
    \mathcal{O}(q^4n^4\sqrt{\epsilon_{inv}/\gamma_q}\log(1/\epsilon_{rect})/\lambda_q)
    .
\end{equation}

The number of oracles and gates that are used in the construction of $\tilde{U}_\Pi$ is as follows:
\begin{align}
    &O^K_q : 
    \mathcal{O}\left(
    \frac{q^4n^6}{(\gamma_q)^2\lambda_q}\log\left(\frac{1}{\epsilon_{rect}}\right)\log\left(\frac{1}{\gamma_q \epsilon_{inv}}\right)\right),\\
    &O^L_{q+1}:\mathcal{O}\left(
    \frac{q^4n^6}{(\gamma_q)^2\lambda_q}\log\left(\frac{1}{\epsilon_{rect}}\right)\log\left(\frac{1}{\gamma_q \epsilon_{inv}}\right)\right),\\
    &{\rm gates}: 
    \tilde{\mathcal{O}}\left(
    \frac{q^4n^8}{(\gamma_q)^2\lambda_q}
    \log\left(\frac{1}{\epsilon_{rect}}\right)\log\left(\frac{1}{\gamma_q \epsilon_{inv}}\right)\right).
\end{align}

\section*{Acknowledgement}
RH thanks Tomoyuki Morimae, Seiichiro Tani, and Yuki Takeuchi for the helpful discussion. RH thanks Tomoyuki Morimae for his comments on the draft. 
{RH thanks anonymous reviewers for their helpful comments to improve the manuscript.}
RH was supported by JSPS KAKENHI Grant Number JP22J11727.

\bibliographystyle{plainnat}
\bibliography{topological}

\appendix

\section{Proofs of Section 2 and 3}
\label{app:proof}

{\it Proof of Theorem \ref{them:1}.} 
Choose any orthonormal basis of $C^{L,K}_{q+1}$ and let $B^{L,K}_{q+1}$ be the corresponding matrix representation of $\partial^{L,K}_{q+1}$. 
Then
    $\Delta^{K,L}_q
    =B^{L,K}_{q+1}\left(B^{L,K}_{q+1}\right)^\mathsf{T}
    +
    \left(B^K_q\right)^\mathsf{T}B^K_q$ using \eqref{eq:matrixrep}.
It follows using Claim \ref{claim:hodge} that 
\begin{equation}
    \beta^{K,L}=\rm{dim}\left(
    \rm{ker}(B^K_q)/\rm{im}(B^{L,K}_{q+1})
    \right)
    =\rm{dim}\left(\rm{ker}(\Delta^{K,L}_q)\right)
    =\rm{nullity}\left(\Delta^{K,L}_q\right)
\end{equation}
with $B^K_qB^{L,K}_{q+1}=0$. \fbox
\\ \\ 

\noindent
{\it Proof of Theorem \ref{theorem:qsvt}.}
This is a slight modification of Corollary 18 of  \cite{gilyen2019quantum}.
Using Corollary 10 of~\cite{gilyen2019quantum}, 
we can find a degree-$d$ polynomial $P\in {\mathbb C}[x]$ s.t. $|\Re[P(x)]-f(x)|\leq \delta$ for all $x\in [-1,1]$ and $P$ satisfies the following conditions:
\begin{itemize}
    \item $P$ has parity-($d$ mod 2),
    \item $\forall x \in [-1,1] : |P(x)|\leq 1$,
    \item $\forall x \in (-\infty,-1]\cup [1,\infty):|P(x)|\geq 1$,
    \item if $d$ is even, then $\forall x\in {\mathbb R}:P(ix)P^*(ix)\geq 1$,
\end{itemize}
with an ${\mathcal{O}(poly(d,\log(1/\delta)))}$-time classical algorithm. 
By Theorem 17 of \cite{gilyen2019quantum},
we can also find a parameter $\Phi=\{\phi_0,\phi_1,...,\phi_d\}\in {\mathbb R}^{d+1}$
in classical ${\mathcal{O}(poly(d,\log(1/\delta)))}$ time
s.t. the parametrized circuit $U_\Phi$ 
satisfies 
\begin{equation}
 P^{({\rm SV})}(\tilde{\Pi} U\Pi)=
 \begin{cases}
 \tilde{\Pi} U_\Phi\Pi\qquad {\rm if}\ d\ {\rm is}\ {\rm odd}, \\
 \Pi U_\Phi\Pi\qquad {\rm if}\ d\ {\rm is}\ {\rm even}.
 \end{cases}
\end{equation}
   Here, $U_\Phi$ is a unitary defined as
\begin{equation}
    U_\Phi\coloneqq
    \begin{cases}
    e^{i\phi_1(2\tilde{\Pi}-I)}U\prod_{j=1}^{(d-1)/2} 
    \left(e^{i\phi_{2j}(2\Pi-I)}U^\dagger
    e^{i\phi_{2j+1}(2\tilde{\Pi}-I)}U
    \right)\ {\rm for\ odd}\ d,\\
    \prod_{j=1}^{d/2} 
    \left(e^{i\phi_{2j-1}(2\Pi-I)}U^\dagger
    e^{i\phi_{2j}(2\tilde{\Pi}-I)}U
    \right)\ {\rm for\ even}\ d.
    \end{cases}
\end{equation}
Each of $e^{i\phi(2\Pi-I)}$ can be implemented using a single ancilla qubit and C$_\Pi$NOT as 
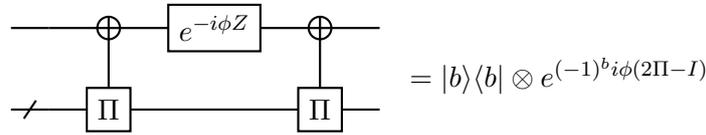
\begin{center}
\begin{quantikz}
&\qw&\targ{1} & \gate{e^{-i\phi Z}} &  \targ{1} &\qw \\
&\qwbundle{}&\gate[wires=1]{\Pi}\vqw{-1} & \qw & \gate[wires=1]{\Pi}\vqw{-1}  &\qw \\
\end{quantikz}
$\quad =|b\rangle\langle b|\otimes e^{(-1)^b i\phi(2\Pi-I)}$ 
\end{center}
for $b\in \{0,1\}$.
We can linearly combine $U_\Phi$ and $U_{-\Phi}$ by applying Hadamard gates to the ancilla qubit as
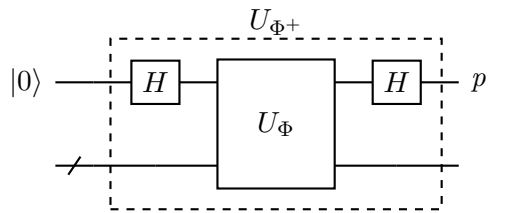
\begin{center}
\begin{quantikz}
\lstick{$\ket{0}$}&\qw & \gate{H}\gategroup[wires=2,steps=3,style={dashed}]{$U_{\Phi^+}$}  & \gate[2]{\quad U_\Phi\quad} & \gate{H} &\qw \rstick{\it p}  \\
& \qwbundle{}& \qw &  &\qw&\qw&
\end{quantikz},
\end{center}
where $p$ means to postselect to $\ket{0}$. 
Define $U_{\Phi^+}$ for odd $d$ and even $d$ as the above figure. 
Let $\Pi'=\tilde{\Pi}$ for odd $d$ and let $\Pi'=\Pi$ for even $d$. 
Then $P^{({\rm SV})}(\tilde{\Pi} U\Pi)=\Pi'U_\Phi\Pi$ and $P^{*({\rm SV})}(\tilde{\Pi} U\Pi)=\Pi'U_{-\Phi}\Pi$, and
\begin{equation}
P_\Re^{({\rm SV})}(\tilde{\Pi} U\Pi)
=\frac{P^{({\rm SV})}(\tilde{\Pi} U\Pi)+P^{*({\rm SV})}(\tilde{\Pi} U\Pi)}{2}
=
(\bra{0}\otimes\Pi')U_{\Phi^+}(\ket{0}\otimes\Pi).
\end{equation}
This satisfies
\begin{equation}
    \left\|f^{(\rm{SV})}(\tilde{\Pi}U\Pi)-P_{\Re}^{(\rm{SV})}(\tilde{\Pi}U\Pi)\right\|\leq \delta.
\end{equation}
It is clear from the construction that 
$U_{\Phi^+}$ 
can be implemented with ${\mathcal O}(d)$ use of $U$, $U^\dagger$, ${\rm C}_\Pi {\rm NOT}$, ${\rm C}_{\tilde{\Pi}} {\rm NOT}$ and single-qubit gates.
\fbox
\\\vspace{0.3cm}

\noindent
{\it Proof of Corollary \ref{corol:qsvtreal}.}
This follows from the construction of $U_{\Phi^+}$ in the proof of Theorem~\ref{theorem:qsvt}.\ \fbox
\\ \vspace{0.3cm}

\noindent
{\it Proof of Theorem \ref{theorem:pseudoinverse}.} 
This is a slight modification of Theorem 41 of \cite{gilyen2019quantum} 
for the block-encoding of the positive semidefinite operator. 
Let $\epsilon'=\frac{\alpha\epsilon}{2}$. 
It is shown in \cite{gilyen2019quantum,martyn2021grand} that there exists an odd real polynomial $f(x)$ that $\frac{\epsilon'}{2\kappa}$-approximates the function $\frac{1}{2\kappa x}$ for $\forall x\in [-1,1]\backslash [-\frac{1}{\kappa},\frac{1}{\kappa}]$, and $|f(x)|\leq 1$ for $\forall x\in [-1,1]$ whose degree is $d = {\mathcal O}(\kappa \log{(\kappa/\epsilon')})$.  ($f(x)$ satisfies $f(0)=0$.)

Let 
$A=\sum_i \lambda_i |\tilde{\psi}_i\rangle\bra{\psi_i}$ be the singular value decomposition of $A$. 
Then 
\begin{equation}
    A/\alpha
    = (\bra{0}^a\otimes I_n)U(\ket{0}^a\otimes I_n)
    =
    \sum_i \lambda_i/\alpha |\tilde{\psi}_i\rangle\bra{\psi_i}.
\end{equation}
By Corollary \ref{corol:qsvtreal},
we can construct $U_{\Phi^+}$ that is a 
QSVT with polynomial $P(x)$ that $\frac{\epsilon'}{2\kappa}$-approximates 
$f(x)$ with $\mathcal{O}(poly(\kappa\log{(\kappa/\epsilon')}))$-time classical algorithm.
Then, because $\lambda_{min}/\alpha > 1/\kappa$, $U_{\Phi^+}$ satisfies
\begin{equation}
    (\bra{0}^{a+1}\otimes I_n)U_{\Phi^+}(\ket{0}^{a+1}\otimes I_n)
    =
    \sum_i P(\lambda_i/\alpha) |\tilde{\psi}_i\rangle\bra{\psi_i}
    \simeq
     \sum_i \frac{\alpha}{2\kappa \lambda_i}|\tilde{\psi}_i\rangle\bra{\psi_i}
    = \frac{\alpha}{2\kappa}A^+.
\end{equation}
Therefore, $(\bra{0}^{a+1}\otimes I_n)U_{\Phi^+}(\ket{0}^{a+1}\otimes I_n)$ is an $(\frac{\epsilon'}{2\kappa}+\frac{\epsilon'}{2\kappa})$-approximation of  $\frac{\alpha}{2\kappa} A^+$.
This means $\frac{2\kappa}{\alpha} (\bra{0}^{\otimes a+1}\otimes I_n)U_{\Phi^+}(\ket{0}^{\otimes a+1}\otimes I_n)$ is $\frac{2}{\alpha}\epsilon'=\epsilon$ approximation of $A^+$.\ \fbox

\section{Explicit construction of the membership oracles}
\label{section:practical}

It is necessary
to be able to construct the membership functions so that we can implement our quantum algorithm efficiently.
Although the number of possible $q$-simplices can be exponential in the size of the data, 
we can implement the membership function if we can efficiently verify whether a simplex is contained at some point of the filtration.
There are many ways of constructing simplicial complexes from the given data, such as the point cloud, the digital image, or the network \cite{otter2017roadmap}.

A $q$-skeleton of a simplicial complex $K$ is a union of $K_p$ for $p=\{0,1,...,q\}$ where $K_p$ is the collection of the $p$-simplices of $K$. 
We can construct the membership functions efficiently for a class of the simplicial complex which is entirely determined by its 1-skeleton 
such as the flag complex, clique complex, Vietoris-Rips complex, and lazy witness complex. 
Besides such simplicial complexes, we are not sure whether we can efficiently implement the membership functions such as the \v{C}ech complex and the Alpha complex. 
In the following, we introduce some constructions of the membership functions.

\paragraph{Vietoris-Rips complex.}
Vietoris-Rips (VR) complex is very commonly used in TDA. This is defined as follows.
Let $S$ be a set of $n$ points in $\mathbb{R}^d$.
Then, the VR complex with parameter $\varepsilon$ denoted by $\mathcal{R}_\varepsilon(S)$ is the set of all $\sigma\in S$ such that the largest distance between any of its points is at most $2\varepsilon$. 
We can construct a VR complex first by computing the graph by connecting any of the vertices of which pairwise distance is less than $2\varepsilon$. 
Then, the VR complex is the clique complex of that graph. 

In order to check whether a set of vertices is contained in the VR complex, we just need to check all of the pairwise distances of the vertices are less than $2\varepsilon$, which can be done in $\mathcal{O}(n^2)$ time in the worst-case. 

It is also possible to implement the membership function for the Vietoris-Rips filtration of the network input \cite{aktas2019persistence}. 
There, an undirected weighted graph is given as input. Then, the filtration is built according to the parameter for the weight of the edges.
The membership verification can be done similarly using the adjacency matrix of the undirected graph with some parameters.

\paragraph{Witness complex.} 
A small subset of the point cloud is used as a `landmark' to reduce the number of simplices. 
Let $S$ be the point set and let $L\subseteq S$ be the landmark set.
Then, the number of possible simplices of the witness complex is $2^{\mathcal{O}|L|}$.
A point $s\in S$ is called a weak $\varepsilon$-witness for $\sigma$ iff 
$d(s,a)\leq d(s,b)$ for all $a\in \sigma$
and $b \in L\ \backslash\ \sigma$ where $d(x,y)$ is the distance points $x$ and $y$.
The weak witness complex at scale $\varepsilon$ is the simplicial complex with vertex $L$ such that all the faces of that simplicial complex $\sigma \subset L$ have a weak $\varepsilon$-witness in $S$. 
The lazy weak witness complex at scale $\varepsilon$ has the same 1-skeleton as the weak witness complex. 
We can efficiently verify membership for the lazy weak witness complex by calculating whether any pair of two points in $L$ has a weak $\varepsilon$-witness.

\section{Construction of the block-encoding of the boundary operator}
\label{sec:appendix:boundaryopeartor}

\subsection{Introduction of $U^K_q$}
As in eq.~\eqref{eq:boundary}, the boundary matrix is defined as 
\begin{equation}
    \partial^K_q([v_0,...,v_q]) \coloneqq 
    \sum_{i=0}^q
    (-1)^i[v_0,...,\hat{v}_i,...,v_q].
\end{equation}
{
In this section, we construct a 
unitary
$U^K_q$ that acts as
\begin{align}
\label{eq:ukq}
U^K_q
&\left(\ket{x}\otimes\ket{0^{\otimes{\lceil \log{(q+1)}\rceil}}}\otimes\ket{0^{\otimes{\lceil \log{n}\rceil}}}\otimes\ket{0^5}\right)\\
&=\frac{1}{\sqrt{{\tilde{n}\tilde{q}}}}\sum_{s=0}^{\tilde{q}-1}\sum_{t=0}^{\tilde{n}-1}
 (-1)^s
 \ket{x\oplus 0^{t-1}10^{n-t}}(H^{\otimes\lceil\log{(q+1)}\rceil}\ket{s})(H^{\otimes\lceil\log{n}\rceil}\ket{t})\\& \hspace{2cm} \otimes
 |h_{q-1}(s),h_n(t),\delta_{s,g_t(x)}\oplus 1,x_t\oplus 1,f^K_q(x)\oplus 1 \rangle
\end{align}
using $\mathcal{\tilde{O}}(n^2)$-number of gates and $\mathcal{O}(1)$-use of $O^K_q$.
Here, 
$h_a(x)=0$ if $x< a$ and $h_a(x)=1$ if $x\geq a$, 
$g_t(x)\coloneqq\sum_{i=0}^{t-1} x_i$ ($x_i$ is the $i$-th bit of $x$), and $\delta_{s,g_t(x)}=1$ if $s=g_t(x)$ and $\delta_{s,g_t(x)}=0$ otherwise.}

\subsection{Verification that $U^K_q$ is the block-encoding of the boundary operator}
{
We show that the unitary $U^K_q$ satisfies the following equations. 
First, we show that for $\forall x\in\bar{S}^K_q\subseteq \{0,1\}^n$, the unitary $U^K_q$ satisfies 
\begin{equation}
\label{eq:be_case1}
    (\bra{0}^a\otimes I_n)U^K_q (\ket{0}^{a}\otimes I_n)
    \ket{x}
    =\frac{1}{{\tilde{n}\tilde{q}}}\partial^K_q \ket{x}
    =
    \frac{1}{{\tilde{n}\tilde{q}}}
    \sum_{i=0}^{q}
    (-1)^i
    \ket{x\oplus 0^{t_i(x)-1}10^{n-t_i(x)}}
\end{equation}
with some $a=\mathcal{O}(\log (n))$. 
Second, we show that for $\forall x\notin\bar{S}^K_q$, 
the unitary $U^K_q$ satisfies 
\begin{equation}
\label{eq:be_case2}
    (\bra{0}^a\otimes I_n)U^K_q (\ket{0}^{a}\otimes I_n)
    \ket{x}
    =0.
\end{equation}
Here, $t_i(x)$ is the index of the $i$-th nonzero bit of $x$, $\tilde{n}\coloneqq 2^{\lceil \log{n}\rceil}$, and $\tilde{q}\coloneqq 2^{\lceil \log{(q+1)}\rceil}$.
Such $U^K_q$ is an $({\tilde{n}\tilde{q}},a,0)$-block-encoding of $\partial^K_q$.

By projecting the fourth register of eq.~\eqref{eq:ukq} to $\ket{0^5}$, we obtain
\begin{align}
\label{eq:motivation}
\frac{\delta_{f^K_q(x),1}}
{\sqrt{{\tilde{n}\tilde{q}}}}&\sum_{s=0}^{q}\sum_{t=0}^{n-1}
 (-1)^s \delta_{s,g_t(x)}\delta_{x_t,1}
\ket{x\oplus 0^{t-1}10^{n-t}}(H^{\otimes\lceil\log{(q+1)}\rceil}\ket{s})(H^{\otimes\lceil\log{n}\rceil}\ket{t})
 \ket{0^5}\\ 
 &=
 \frac{\delta_{f^K_q(x),1}}{\sqrt{{\tilde{n}\tilde{q}}}}\sum_{s=0}^{q}
 (-1)^s 
\ket{x\oplus 0^{t_s(x)-1}10^{n-t_s(x)}}(H^{\otimes\lceil\log{(q+1)}\rceil}\ket{s})(H^{\otimes\lceil\log{n}\rceil}\ket{t})
 \ket{0^5},
\end{align}
where we denote by $t_s(x)$ the $t$ s.t. $g_t(x)=s$ and $x_t=1$. 
The $t_s(x)$ represents the index of the $s$-th non-zero bit of $x$.
Then, by further projecting the second and the third registers to $\ket{0}$, we obtain
\begin{align}
&\left( I_n\otimes\bra{0^{{\lceil \log{(q+1)}\rceil}}}\otimes\bra{0^{{\lceil \log{n}\rceil}}}\otimes\bra{0^5}\right)
U^K_q
\left(
\ket{x}\otimes\ket{0^{{\lceil \log{(q+1)}\rceil}}}\otimes\ket{0^{{\lceil \log{n}\rceil}}}\otimes\ket{0^5}\right)  \\
&
=\frac{\delta_{f^K_q(x),1}}{{\tilde{n}\tilde{q}}}\sum_{s'=0}^{\tilde{q}-1}\sum_{t'=0}^{\tilde{n}-1} \sum_{s=0}^{q}
 (-1)^s
\ket{x\oplus 0^{t_s(x)-1}10^{n-t_s(x)}}\langle s'\ket{s}\langle {t'}\ket{t_s(x)}\\
&= 
\frac{\delta_{f^K_q(x),1}}{{\tilde{n}\tilde{q}}} \sum_{s=0}^{q}
 (-1)^s
\ket{x\oplus 0^{t_s(x)-1}10^{n-t_s(x)}}.
\end{align}
Therefore, the unitary $U^K_q$ satisfies eq.~\eqref{eq:be_case1} and eq.~\eqref{eq:be_case2}.
}

\subsection{Construction of $U^K_q$}
We construct the unitary $U^K_q$ in the following 5 steps.

\begin{enumerate}

\item 
Generate the following state:
\begin{equation}
\label{eq:step0}
\frac{1}{\sqrt{{\tilde{n}\tilde{q}}}}\sum_{s=0}^{\tilde{q}-1}\sum_{t=0}^{\tilde{n}-1}\ket{x}\ket{s}\ket{t}\ket{0^5}.
\end{equation}
This state can be generated by applying $H^{\otimes\lceil\log{(q+1)}\rceil}$ and $H^{\otimes\lceil\log{n}\rceil}$ to the second and the third registers of $\ket{x}\otimes\ket{0^{\otimes{\lceil \log{(q+1)}\rceil}}}\otimes\ket{0^{\otimes{\lceil \log{n}\rceil}}}\otimes\ket{0^5}$, respectively.

\item
Generate the following state:
\begin{equation}
\label{eq:step2}
\frac{1}{\sqrt{{\tilde{n}\tilde{q}}}}\sum_{s=0}^{\tilde{q}-1}\sum_{t=0}^{\tilde{n}-1}\ket{x}\ket{s}\ket{t}|h_{q-1}(s),h_n(t),\delta_{s,g_t(x)}\oplus 1,x_t\oplus 1,f^K_q(x)\oplus 1\rangle,
\end{equation}
where $h_a(x)=0$ if $x< a$ and $h_a(x)=1$ if $x\geq a$, 
$g_t(x)=\sum_{i=0}^{t-1} x_i$ ($x_i$ is the $i$-th bit of $x$), and $\delta_{s,g_t(x)}=1$ if $s=g_t(x)$ and $\delta_{s,g_t(x)}=0$ otherwise. 
The computed functions $|h_{q-1}(s),h_n(t),\delta_{s,g_t(x)}\oplus 1,x_t\oplus 1,f^K_q(x)\oplus 1\rangle$ in the fourth registers are projected to $\ket{0^5}$ as 
in eq.~\eqref{eq:motivation}. By this projection, we can only leave the sums of $s$ and $t$ which are needed as the boundary operator, which is the motivation for computing these functions.

The generation of the quantum state of eq.~\eqref{eq:step2} can be done as follows.

\begin{itemize}

\item
We first compute $h_{q-1}(s)$ in the first qubit of the fourth register of eq.~\eqref{eq:step0}.
For computing $h_{q-1}(s)$, we first prepare $\ket{q}$ in an ancilla register and compute whether $s\leq q-1$ or $s>q$ using a quantum circuit for subtraction with $\mathcal{O}(\log{n})$ gates~\cite{hayakawa2019fine}. Finally, initialize the ancilla register. 

\item
Similarly, compute $h_n(t)$ in the second qubit of the fourth register.

\item Compute $g_t(x)$ in ancilla qubits as \[
\frac{1}{\sqrt{{\tilde{n}\tilde{q}}}}\sum_{s=0}^{\tilde{q}-1}\sum_{t=0}^{\tilde{n}-1}\ket{x}\ket{s}\ket{t}\ket{h_{q-1}(s),h_n(t),0^3}\ket{g_t(x)}.
\]
This can be done using the quantum circuit for addition~\cite{cuccaro2004new}, which uses another single ancilla qubit and $\mathcal{\tilde{O}}(n^2)$-gates.
\item Flip the first qubit of the fourth register iff $s\neq g_t(x)$: 
\[
\frac{1}{\sqrt{{\tilde{n}\tilde{q}}}}\sum_{s=0}^{\tilde{q}-1}\sum_{t=0}^{\tilde{n}-1}\ket{x}\ket{s}\ket{t}\ket{h_{q-1}(s),h_n(t),\delta_{s,g_t(x)}\oplus 1,0^2}\ket{g_t(x)}.
\]
\item Initialize the ancilla register: 
\[
\frac{1}{\sqrt{{\tilde{n}\tilde{q}}}}\sum_{s=0}^{\tilde{q}-1}\sum_{t=0}^{\tilde{n}-1}\ket{x}\ket{s}\ket{t}\ket{h_{q-1}(s),h_n(t),\delta_{s,g_t(x)}\oplus 1,0^2}\ket{0...0}.
\]

\item
Copy the $t$-th bit of $x$ to the fourth qubit of the fourth register and also apply $X$ gate to the same qubit:
\begin{equation}
\frac{1}{\sqrt{{\tilde{n}\tilde{q}}}}\sum_{s=0}^{\tilde{q}-1}\sum_{t=0}^{\tilde{n}-1}\ket{x}\ket{s}\ket{t}|h_{q-1}(s),h_n(t),\delta_{s,g_t(x)}\oplus 1,x_t\oplus 1,0\rangle.
\end{equation}

\item Apply the membership oracle $O^K_q$ to the first register and the fifth qubit of the fourth register, and also apply $X$-gate to the fifth qubit of the fourth register:
\begin{equation}
\frac{1}{\sqrt{{\tilde{n}\tilde{q}}}}\sum_{s=0}^{\tilde{q}-1}\sum_{t=0}^{\tilde{n}-1}\ket{x}\ket{s}\ket{t}|h_{q-1}(s),h_n(t),\delta_{s,g_t(x)}\oplus 1,x_t\oplus 1,f^K_q(x)\oplus 1\rangle.
\end{equation}
\end{itemize}

\item
Apply a $Z$-gate to the first qubit of the second register:
\begin{equation}
 \frac{1}{\sqrt{{\tilde{n}\tilde{q}}}}\sum_{s=0}^{\tilde{q}-1}\sum_{t=0}^{\tilde{n}-1}
 (-1)^s
 \ket{x}\ket{s}\ket{t}
 |h_{q-1}(s),h_n(t),\delta_{s,g_t(x)}\oplus 1,x_t\oplus 1,f^K_q(x)\oplus 1 \rangle.
\end{equation}

\item
Flip the $t$-th bit of the first register:
\begin{equation}
\label{eq:step6}
 \frac{1}{\sqrt{{\tilde{n}\tilde{q}}}}\sum_{s=0}^{\tilde{q}-1}\sum_{t=0}^{\tilde{n}-1}
 (-1)^s
 \ket{x\oplus 0^{t-1}10^{n-t}}\ket{s}\ket{t}
 |h_{q-1}(s),h_n(t),\delta_{s,g_t(x)}\oplus 1,x_t\oplus 1,f^K_q(x)\oplus 1 \rangle.
\end{equation}

\item Apply $H^{\otimes\lceil\log{(q+1)}\rceil}$ and $H^{\otimes\lceil\log{n}\rceil}$ to the second and the third registers of eq.~\eqref{eq:step6} to obtain 
\begin{align}
\label{ep:beforepostselect}
\frac{1}{\sqrt{{\tilde{n}\tilde{q}}}}\sum_{s=0}^{\tilde{q}-1}\sum_{t=0}^{\tilde{n}-1}
 (-1)^s
 \ket{x\oplus 0^{t-1}10^{n-t}}&(H^{\otimes\lceil\log{(q+1)}\rceil}\ket{s})(H^{\otimes\lceil\log{n}\rceil}\ket{t})\\&
 \otimes|h_{q-1}(s),h_n(t),\delta_{s,g_t(x)}\oplus 1,x_t\oplus 1,f^K_q(x)\oplus 1 \rangle.
\end{align}

\end{enumerate}
We have implemented the unitary of eq.~\eqref{eq:ukq} in the above 5 steps. We have used $\mathcal{\tilde{O}}(n^2)$-number of gates and $\mathcal{O}(1)$-number of $O^K_q$ in this construction.

\end{document}